\documentclass[aps,pra,amssymb,onecolumn,showpacs,a4paper,superscriptaddress]{revtex4-1}

\usepackage{amsmath,amssymb,epsfig,graphicx,color,framed}
\usepackage{hyperref}
\usepackage{amsfonts} 
\usepackage{amsmath}
\usepackage{amssymb}
\usepackage{graphicx}
\usepackage{hyperref}
\usepackage{color}
\usepackage{natbib}  
\usepackage{bm} 
\usepackage{isomath} 
\usepackage{braket}
\usepackage{booktabs} 

\newcommand{\vect}[1]{\vectorsym{#1}} 

\begin{document}

\title{Stochastic resonance in time-delayed bistable active particles} 

\author{Antonio A. Valido}
\email{antonio.valido@ulpgc.es}
\affiliation{iUNAT, Departamento de Física, Universidad de Las Palmas de Gran Canaria, 35017 Las Palmas de Gran Canaria, Spain}

\author{Mattia Coccolo}
\affiliation{Nonlinear Dynamics, Chaos and Complex Systems Group, Departamento de F\'{i}sica,
Universidad Rey Juan Carlos, Tulip\'{a}n s/n, 28933 M\'{o}stoles, Madrid, Spain}

\author{Miguel A.F. Sanju\'{a}n}
\affiliation{Nonlinear Dynamics, Chaos and Complex Systems Group, Departamento de F\'{i}sica,
Universidad Rey Juan Carlos, Tulip\'{a}n s/n, 28933 M\'{o}stoles, Madrid, Spain}

\affiliation{Royal Academy of Sciences of Spain, Valverde 22, 28004 Madrid, Spain}

\date{\today}

\begin{abstract}
We investigate analytically and numerically stochastic resonance in an underdamped active (Ornstein--Uhlenbeck) particle confined in a bistable potential with time-delayed feedback and subjected to an additional (non-thermal) white-noise source. Restricting the analysis to the active-noise-dominated bistable regime, with sufficiently short delay and active-noise correlation times, we develop a perturbative theory that yields analytical expressions for both the escape rate and the spectral power amplification within the framework of linear-response theory, without invoking detailed balance. The analytical predictions are in qualitative agreement with direct numerical simulations throughout the perturbative regime. We show that a moderate increase in the active-noise correlation time can enhance the stochastic-resonance response, in marked contrast to previous results for overdamped passive systems driven by colored noise. Conversely, the addition of white noise weakens the resonant response and shifts its optimum toward lower active-noise intensities. Time-delayed feedback further alters the matching condition between the escape rate and the driving frequency while preserving the resonant mechanism. Beyond the perturbative regime, we find that finite-memory effects progressively broaden the resonance profile and substantially decrease its amplitude. Overall, these results provide new insight into the combined role of inertia, activity, environmental fluctuations, and time delayed feedback in stochastic resonance in nonequilibrium systems.

\end{abstract}

\maketitle

\section{Introduction}

Stochastic resonance (SR) is one of the most remarkable manifestations of constructive noise in nonlinear systems \cite{anishchenko19991,rajasekar20161}, whereby a weak periodic signal is optimally amplified by random fluctuations. Since its discovery in the context of climate dynamics \cite{benzi19811}, SR has been observed in a broad range of physical \cite{gammaitoni19981}, biological \cite{hanggi20021}, and engineering systems, including mechanical beams \cite{tweten20141}, neuronal dynamics \cite{nozaki19991}, and nanoscale devices \cite{paneru20231}. More recently, the rapid development of active-matter physics has motivated the extension of SR concepts to intrinsically nonequilibrium systems \cite{wang20251,zeng20151,jia20261}. In particular, the active Ornstein--Uhlenbeck particle (AOUP) model has emerged as a paradigmatic framework for describing active fluctuations \cite{fodor20161}, accounting for the finite persistence of self-propelled motion through a characteristic correlation time \cite{maggi20171,dabelow20211,baek20231,bonilla20191,martin20211,woillez20201}. In many active systems, these persistent fluctuations coexist with delayed interactions and additional thermal or environmental noise, introducing multiple competing time scales whose interplay may govern the emergence and characteristics of SR.

The influence of memory effects on stochastic resonance has been extensively investigated in passive systems, i.e., in the absence of self-propulsion, through both temporally correlated noise and delayed feedback. Early studies demonstrated that colored noise can profoundly modify the resonant response by shifting the optimal noise intensity \cite{hanggi19931,liu20251,kang20031,gammaitoni19892}, and even inducing a transition from a single to a double resonance peak in overdamped systems \cite{jia20011,luo20031}. In particular, experimental studies on sensory neurons further revealed that the spectral properties of additive colored noise strongly affect signal amplification \cite{nozaki19991}, establishing temporal correlations as a key ingredient governing SR. Another important source of memory is time-delayed feedback \cite{frank20051,coccolo20211,cantisan20201}. In bistable systems, delayed interactions modify the escape dynamics by introducing an additional characteristic time scale that affects the synchronization between noise-induced barrier crossings and the external periodic forcing. In particular, it has been shown to enhance \cite{gu20121,mei20091} or degrade \cite{du20111,gao20161} the SR response depending on the strength of the time-delayed feedback and the damping rate. We also note that the escape dynamics of the AOUP model has been previously studied for the situation of a harmonic trap \cite{wexler20201,woillez20201}, a bistable (Duffing) potential \cite{sharma20171}, or for more generic confining potentials in the overdamped scenario \cite{woillez20201}. These findings indicate that the interplay between temporal correlations of stochastic fluctuations and time delay can dramatically alter the resonant response, although its behavior remains strongly model dependent.

Despite these advances, the combined effects of inertia, active fluctuations, time delay and (non-thermal) white noise on SR remain largely unexplored. While the dynamics of an active time-delayed bistable particle has recently been investigated in Ref.~\cite{valido20231}, previous studies have mainly focused either on SR in passive systems driven by thermal white or colored noise \cite{gammaitoni19891,kang20031,madureira19951,hanggi19941,jung19932,gammaitoni19892,hanggi19931,bartussek19951,dykman19931}, or on noise-induced escape in active systems without time-delayed feedback \cite{caprini20191,baek20231,woillez20201}. Consequently, a theoretical description of SR in underdamped active systems simultaneously subjected to a time-delayed feedback and a (non-thermal) white noise is still lacking.

In this work, we address this problem by considering an underdamped AOUP evolving in a delayed bistable (Duffing) potential under weak periodic forcing and subject to both active and non-thermal environmental fluctuations. The delayed feedback modifies the effective potential barrier, whereas the active process introduces persistent nonequilibrium fluctuations characterized by a finite correlation time. To gain analytical insight, we develop a perturbative treatment valid for sufficiently weak noise intensities and short delay and active-noise correlation times. Specifically, we combine a Wentzel--Kramers--Brillouin (WKB) approximation \cite{klosek19861,klosek19881} for the stationary solution of the Kramers equation with linear-response theory \cite{jung19931,jung19911,jung19932}. Importantly, this approach retains the essential nonequilibrium character of the nonlinear active dynamics without imposing detailed balance, allowing us to derive analytical expressions for both the escape rate and the spectral power amplification (SPA) \cite{jung19911}.

The analytical predictions are complemented by direct numerical simulations, allowing us to assess the robustness of the perturbative theory and to explore the dynamics beyond its formal range of validity. Within the perturbative regime, the numerical results support the predicted location of the stochastic-resonance region and the corresponding escape-rate matching condition. However, the absolute resonance amplitude is more sensitive to finite-ensemble effects, preventing a fully quantitative validation of the analytical predictions for the peak height. In particular, increasing the active-noise correlation time may enhance the resonant response, in marked contrast to previous findings for overdamped systems driven by colored noise. Environmental white noise, on the other hand, weakens the resonance and shifts the optimal active-noise intensity toward lower values, whereas time-delayed feedback primarily modifies the matching condition between the noise-induced escape rate and the driving frequency while preserving the underlying resonant mechanism. Extending the analysis beyond the perturbative regime reveals that stronger finite-memory effects progressively broaden the resonance profile and reduce its amplitude. Overall, these findings elucidate how inertia, active persistence, environmental fluctuations, and delayed feedback jointly shape stochastic resonance in nonequilibrium systems, providing a theoretical basis for controlling noise-assisted signal amplification and transport in active systems with memory.

The remainder of the paper is organized as follows. In Sec.~\ref{SecMMT}, we introduce the time-delayed bistable active model together with the theoretical framework employed throughout the paper, paying particular attention to the spectral properties of the asymptotic time probability density function. Section~\ref{SecARSTD} is devoted to the perturbative analytical treatment, where the main theoretical results for the escape rate and the SPA are obtained. In Sec.~\ref{SecNMS}, the analytical predictions are validated against direct numerical simulations, and the analysis is extended to the small-to-intermediate delay and active-noise correlation-time regime, where finite-memory effects become increasingly important. Finally, Sec.~\ref{SecOC} presents the conclusions and an outlook on future developments.

\section{Model and methods}\label{SecMMT}

In this section, we specify the system and parameter regime considered throughout the analysis and characterize its SR response through a power-spectrum analysis, introducing the SPA together with the methodology used to compute it. 

\subsection{Time-delayed bistable active dynamics}
We consider the dynamics of an AOUP under the simultaneous action of a time-delayed double-well potential and a periodic forcing. The stochastic equation that governs the underdamped dynamics for its spatial coordinate $X(t)$ reads
\begin{align}
    \frac{d X(t)}{dt}&=V(t), \label{EqDON0} \\
    m\frac{d V(t)}{dt}&=-\mu V(t)-\frac{\partial U_{\tau}(X,t)}{\partial X}+F\cos(\Omega t)+\sqrt{2D_{wn}}\Xi_{wn}(t)+\sqrt{2D_{an}}\xi_{an}(t), \label{EqDON} \\
    \frac{d \xi_{an}(t)}{dt}&=\frac{1}{\tau_{an}}\Big[(\xi_{an}^{\infty} -\xi_{an}(t))+\sqrt{2C_{an}'} \Xi_{an}(t)\Big], \label{EqCOU}
\end{align}
where $\Xi_{wn}(t)$ and $\xi_{an}(t)$ represent, respectively, the environmental and self-propulsion forces modeled by Gaussian white and OU noises. Here, $D_{wn}$, $D_{an}$, and $\mu$ denote the intensities of the white noise and the OU noise (also called active noise \cite{fodor20161,martin20211,doering19871}), and the friction coefficient; whereas $\Omega$ and $F$ represent the (angular) frequency and the amplitude of the time-periodic force.  

We introduce the active noise $\xi_{an}$ as an additional state variable, rendering the extended process $(X,V,\xi_{an})$ Markovian upon a perturbative treatment of the time-delayed dynamics. We consider that $(X,V)$ is initially uncorrelated with $\xi_{an}$ and that the white noises $\Xi_{wn}$ and $\Xi_{an}$ are statistically independent, i.e., $\langle\Xi_{wn}(t)\Xi_{an}(t')\rangle=0$. Moreover, for $0<t'<t$, we have assumed vanishing mean-values $\langle \Xi_{wn}(t)\rangle_{wn}=\langle \Xi_{an}(t)\rangle_{an}=\langle \xi_{an}(t)\rangle_{an}=0$, whereas the covariance functions are given by
\begin{align}
    C_{wn}(t-t')&=C_{wn}'\delta(t-t'), \label{EQWN} \\
    C_{an}'(t-t')&=C_{an}'\delta(t-t'), \label{EQWNAN} 
\end{align}
which yields for the active noise
\begin{equation}
 C_{an}(t-t')=\frac{C_{an}'}{\tau_{an}}e^{-\frac{(t-t')}{\tau_{an}}}, \label{EQOUN}
\end{equation}
where $\delta$ denotes the Dirac-delta function, and $\tau_{an}$ denotes the active-noise correlation time. The quantities $C_{wn}$ and $C_{an}$ are the amplitude of the white noise and the strength of the self-propulsion, respectively. Here, we study the scenario in which the integrated active-noise intensity, $\int_{-\infty}^{\infty} C_{\mathrm{an}}(t),dt$, is kept fixed as $\tau_{\mathrm{an}}$ varies, while the white-noise contribution remains perturbative. This defines the active-noise-dominated bistable regime, enforced through the following condition \cite{valido20231}
\begin{equation}
C_{wn}'=\frac{D_{an}}{\tau_{an}}, \qquad C_{an}'=1.
\label{EqSN}
\end{equation}
Unlike studies of thermally activated escape over fluctuating barriers, where thermal and colored-noise contributions are of comparable strength \cite{reimann19981,bartussek19951}, in our work the active noise dominates interwell transitions and the white noise acts only as a weak perturbation. Accordingly, the noise-induced dynamics of SR will be characterized mainly by $D_{an}$ and $\tau_{an}$. Furthermore, in previous studies of SR \cite{madureira19951,hanggi19941,jung19911,jung19932}, $\xi_{an}$ is typically treated as thermal colored noise, with $D_{an}=\mu k_BT/2$ fixed by the fluctuation--dissipation relation. Here, instead, $D_{an}$ is an independent nonequilibrium parameter, unrelated to $\mu$, so neither fluctuation--dissipation nor detailed balance is assumed.

Additionally, $U_{\tau}$ denotes the time-delayed double-well potential \cite{coccolo20211,cantisan20201,valido20231}, i.e.,
\begin{equation}
    U_{\tau}(X,t)=\frac{\beta}{4}X^4(t)-\frac{\alpha}{2}X^2(t)+\gamma X(t)X(t-\tau),
    \label{PtFunc}
\end{equation}
where the last term in the right hand represents the time-delayed feedback of strength $\gamma$ and time delay $\tau$, whilst $\alpha$ and $\beta$ denote the linear and nonlinear coupling coefficients. For the sake of clarity, we shall consider a constant historical time delay function~\cite{cantisan20201} which cancels for $t<\tau$, i.e., $X(t-\tau) =0$ when $t<\tau$. As stated in the Introduction, we focus on the bistable regime in which the noiseless, undriven system has an unstable fixed point at the origin and two stable equilibria at the potential minima, i.e.,
\begin{equation}\label{eq:2}
X^*_{0}=0, \quad X^*_{1,2}=\pm\sqrt{\frac{\alpha-\gamma}{\beta}},
\nonumber
\end{equation}
for $\alpha,\beta>0$ and $\alpha>\gamma$. The wells are separated by a barrier of height $\Delta U=(\alpha-\gamma)^2/(4\beta)$, modulated by the delayed feedback. As shown in Ref.~\cite{valido20231}, the delay time $\tau$ and feedback strength $\gamma$ strongly affect the stability and long-time dynamics: sufficiently large $\tau$ can sustain aperiodic motion, while strong noise and forcing may induce approximately periodic interwell motion associated with a limit-cycle-like trajectory spanning both wells. Here, we restrict the analysis to the regime in which $X_0^*$ remains unstable and $X_{1,2}^*$ locally stable as $\tau$ varies. We carry out a linear stability analysis in App.~\ref{SecLSA} and show that this preserves the two metastable states required for SR \cite{hanggi20001,gammaitoni19981}. We further consider the weak-noise regime, where interwell transitions in the absence of forcing are rare and much slower than intrawell relaxation. This time-scale separation requires sufficiently small noise intensities and active-noise correlation time \cite{reimann19981,klosek19861,klosek19881,hanggi19941,fox19881}. The corresponding validity conditions are discussed in Sec.~\ref{SecARSTD} (see Eqs.~(\ref{PerWN_1})--(\ref{PerWN_3})). Unlike most previous studies \cite{hanggi19931,klosek19881,jung19911,jung19931,madureira19951,klosek19861}, we also consider moderate-to-strong damping \cite{hanggi19851}, covering both the underdamped, $(\alpha-\gamma)m/\mu^2\sim1$, and overdamped, $(\alpha-\gamma)m/\mu^2\ll1$, regimes.

Within the previously defined regime of validity, the SPA can be evaluated analytically by means of linear-response theory \cite{jung19931}. Concretely, this approach leads to an analytical expression for the resonant response, see Eq.~(\ref{etaLRTapp}), in the small time delay and active-noise correlation time scenario, and allows us to determine the optimal active-noise strength associated with SR, see Eq.~(\ref{EqDOPT}). We shall show in Sec. \ref{SecARSTD} that the location and amplitude of this maximum are not universal, but depend on the time delay $\tau$, the white-noise intensity $D_{\mathrm{wn}}$, and the active-noise correlation time $\tau_{\mathrm{an}}$, since these parameters modify the effective escape rate and, consequently, the matching condition between interwell transitions and the external forcing. Finally, the validity of the analytical prediction is assessed by comparison with numerical simulations of the full stochastic bistable dynamics.

\subsection{Spectral power amplification}\label{SecPSA}

The SPA of the stochastic process $X(t)$ at the driving frequency $\Omega$ \cite{jung19931,jung19932,jung19911}, defined as
\begin{equation}
    \eta(\tau,\tau_{an}, D_{wn},D_{an})=4\bigg(\frac{m|\tilde{X}_{1}|}{F}\bigg)^2.
    \label{SPAE}
\end{equation}
A non-monotonic dependence of the SPA on the active-noise intensity, featuring a maximum at an intermediate value, is taken as the spectral signature of SR \cite{gammaitoni19981,anishchenko19991,jung19911}.
Here, $\tilde{X}_{1}$ is the Fourier coefficient of the first moment of the time-periodic spatial coordinate $X(t)$, evaluated in the asymptotic regime (i.e. $t_{0}\rightarrow-\infty$) evaluated at $\omega=\Omega$. The latter, denoted by $\langle X(t)\rangle_{as}$, is given by
\begin{equation}
    \langle X(t)\rangle_{as}=\int_{-\infty}^{\infty}\int_{-\infty}^{\infty}xP_{as}(x,v,t)dxdv,
    \label{MVXASymp}
\end{equation}
where $P_{as}(x,v,t)$ denotes the time-dependent asymptotic one-time probability density function of the reduced system $(X,V)$. This distribution follows 
from the kinetic equations associated with the underlying stochastic 
dynamics (\ref{EqDON0})--(\ref{EqCOU}). 

Obtaining a closed-form expression for $P_{as}(x,v,t)$ is challenging because of both the nonlinear potential and the time-dependent periodic forcing \cite{madureira19951,caprini20191,woillez20201,kosek19881}. To circumvent this difficulty, we approximate $\tilde{X}_1$ using linear-response theory \cite{hanggi19821,hanggi19931,jung19931,jung19932},
\begin{equation}
    |\tilde{X}_{1}|=\frac{F}{2m}|\chi(\Omega)|,
    \label{FTRFS1} 
\end{equation}
where we have introduced the Fourier transform of the susceptibility, denoted by $\chi(\omega)$. Hence, we show in App. \ref{App4} that the susceptibility can be further approximated by invoking a non-equilibrium steady-state fluctuation-response relation and the so-called long-time approximation, i.e.,
\begin{equation}
    \chi(\omega)
    =\frac{2c_{\tau}\langle x^2\rangle_{st}\lambda_{ac}}
    {\lambda_{ac}+i\omega},
    \label{EqSuc}
\end{equation}
where $\lambda_{ac}$ and $\langle x^2\rangle_{st}$  are the lowest nonzero eigenvalue of the kinetic equation and the stationary mean-square displacement in the absence of periodic driving, respectively, while $c_{\tau}$ is an auxiliary coefficient. In the weak-noise regime, $\lambda_{ac}$ can be well approximated by the escape rate $k_{ac}$ in the absence of the periodic driving \cite{jung19931,bartussek19951,anishchenko19991}. Owing to the symmetry of the bistable potential, the fundamental nonzero eigenvalue of the relaxation spectrum is identified as $\lambda_{ac}=2k_{ac}$. We then obtain an analytical expression for the SPA in the small time-delay and short active-noise correlation-time regime by evaluating $k_{ac}$ within Kramers procedure using the flux-over-population method \cite{kramers19401}. The corresponding validity conditions are specified in the following section.

Additionally, we investigate the SR phenomenon beyond the aforementioned approximations by performing direct numerical simulations of the stochastic time-delayed differential equations. The dynamics is integrated over a sufficiently long time,
\(
T_{\infty}\geq100\,2\pi/\Omega,
\)
to ensure that the system reaches its asymptotic periodic state. The numerical integration is carried out using the explicit stabilized Runge--Kutta (S-ROCK) scheme \cite{komori20191} implemented in \textsc{Matlab}. A fixed time step of $dt=10^{-3}$ is employed throughout the simulations, while the active noise is propagated using its exact discrete-time update, thereby preserving its prescribed correlation time. We sample $\xi_{an}$ from its stationary distribution, ensuring stationary noise correlations. The asymptotic mean response,
\(
\langle X(t)\rangle_{as},
\)
is obtained by averaging over an ensemble of $N_{\rm sim}$ independent trajectories with identical initial conditions but different realizations of both the white and active noises. The corresponding Fourier coefficients are computed as
\begin{equation}
\langle X (t)\rangle_{as}=
\sum_{k=-N_F}^{N_F}
\tilde{X}_k
e^{i(\Omega dt+\bar{\phi}_{k}) k},
\end{equation}
where $N_F$ is chosen sufficiently large to accurately resolve the asymptotic periodic response~\cite{anishchenko20071}. We have verified the convergence of the numerical results with respect to both the time step and the ensemble size for $N_{\rm sim}\gtrsim100, 200,400,800$; finding that the position of the resonant region is much more robust than its absolute amplitude (see Fig. \ref{fig:CAS} in App. \ref{App5}). 

The eigenvalue $\lambda_{ac}$ is also estimated numerically from the mean first-passage time (MFPT), denoted by $T_{MFPT}$. For this purpose, the external periodic forcing is switched off, and $N_{\rm sim}$ independent trajectories are initialized at the left stable equilibrium $X_{1}^{*}$. We define the escape time $T_{\rm esc}$ as the first-passage time to the barrier top located at $X_{0}^{*}=0$. In the presence of delayed feedback, the origin remains the unstable equilibrium of the deterministic system, as shown in App. \ref{SecLSA}. For sufficiently large $N_{\rm sim}$, its ensemble average approaches the MFPT, i.e., $\langle T_{\rm esc}\rangle \rightarrow T_{MFPT}$. In the rare-hopping regime, the inverse of the escape rate coincides with the MFPT \cite{jung19931,naeh19901,reimann19991}, i.e., $T_{MFPT}=k_{ac}^{-1}$ \cite{hanggi19901}. The escape rate is then estimated as
$k_{ac}=\frac{1}{\langle T_{\rm esc}\rangle}$ \cite{reimann19991}, where the average is taken over all trajectories that escape before a maximum integration time $T_{\max}$. As a convergence criterion, the fraction of trajectories escaping before $T_{\max}$ is monitored throughout the simulations. Parameter sets for which fewer than $80\%$ of the trajectories escape are discarded from the quantitative analysis of $\lambda_{ac}$.

\section{Analytical results: small time-delay and short active-noise correlation time}
\label{SecARSTD}

As anticipated in the previous section, to determine the susceptibility, we first need to compute the escape rate, which in turn requires an analytical expression for the stationary distribution of the kinetic equation in the absence of periodic driving. To obtain the latter, we restrict the analysis to the scenario in which the delay-induced memory effects remain weak compared with the intrinsic relaxation dynamics, that is
\begin{equation}
\frac{\gamma\tau}{[m(\alpha-\gamma)]^{1/2}}
\sim \frac{\gamma\tau}{\mu}\ll 1,
\label{PerTD}
\end{equation}
which is widely used in time-delayed stochastic systems \cite{frank20051,gu20121,kumaran20221,mei20091,gao20161,qiu20231}. As shown in App.~\ref{App1}, this condition allows us to recast the kinetic equation into a (closed) time-dependent Fokker--Planck (FP) equation in the enlarged phase space $(X,V,\xi_{an})$; see Eqs.~(\ref{AppFKEGI})--(\ref{FPEq}). Exact stationary solutions can be obtained by quadratures when the drift and diffusion coefficients satisfy specific constraints, such as detailed balance \cite{risken19841}. Such conditions, however, are generally incompatible with the nonequilibrium nature of active matter. We therefore determine the stationary probability density of the FP equation using a singular perturbation method \cite{klosek19861,klosek19881}, analogous to a WKB approximation. Remarkably, at the perturbative order considered here, the effective FP dynamics yields an equilibrium-like stationary distribution without requiring detailed balance. This analytical expression is presented in the following section.

\subsection{Stationary probability density function, escape rate and susceptibility}\label{SubsectSSPDF}

To apply the singular perturbation method, we restrict the analysis to the weak-noise regime, where the noise strength is sufficiently small compared with the potential barrier height to serve as a perturbative parameter, i.e.,
\begin{equation}
    \frac{D_{an}}{(m(\alpha-\gamma))^{\frac{1}{2}}\Delta U} \ll 1 .
    \label{PerWN_1}
\end{equation}
Similar weak-noise conditions are widely employed in studies of stochastic resonance and escape-rate problems \cite{hanggi19901,gammaitoni19981} and are fully consistent with the SR regime considered here. This allows the steady-state probability density function, denoted by $P_{st}(x,v,\xi)$, to be approximated by a WKB form \cite{kosek19881}, i.e.,
\begin{equation}
    P_{st}(x,v,\xi)\approx Z^{-1}e^{-\Phi_{wkb}(x,v)},
    \label{AppWKBASS}
\end{equation}
where $Z$ is the normalizing constant and $\Phi_{wkb}$ is the so-called quasipotential distribution. Yet, it is difficult to obtain a closed-form expression for the latter due to the underlying nonlinear and non-Markovian nature of the dynamics (recall that the correlation time of the active noise (\ref{EqCOU}) is finite). It is therefore necessary to further restrict our analytical study to the sufficiently small correlation time $\tau_{an}$ and white-noise strength $D_{wn}$. Concretely, we further demand the following conditions
\begin{align}
  0<\frac{D_{an}D_{wn}}{(m(\alpha-\gamma))^{\frac{1}{2}}\Delta U\tau_{an}}&\ll 1, \label{PerWN_2}\\
 0<\frac{\tau_{an}(\alpha-\gamma)^{\frac{1}{2}}}{m}&\ll 1.\label{PerWN_3}
\end{align}
The inequality (\ref{PerWN_2}) ensures the perturbative character of the white noise compared with the active noise, whereas the condition in Eq.~(\ref{PerWN_3}) governs the quasi-Markovian regime of interest (recall that the correlation function $\langle\xi_{an}(t)\xi_{an}(t')\rangle$ approaches a delta function in the limit $\tau_{an}(\alpha-\gamma)^{\frac{1}{2}}/m\rightarrow0$). Specifically, the latter establishes that the memory time scale of the active noise is much shorter than the deterministic relaxation time scale imposed by the quartic potential. Under these conditions we follow and extend the prescription of Ref.~\cite{kosek19881} to obtain an approximate expression for $\Phi_{wkb}$ to leading order, which reads
\begin{align}
    \Phi_{wkb}(x,v)&=c_{\tau}v^2+d_{\tau}U_{0}(x),
    \label{QPDSD}
\end{align}
where we have introduced
\begin{align}
     c_{\tau}&=\frac{\mu }{4 D_{an}}\bigg[ma_{\tau}+\mu \tau_{an}b_{\tau}^2 -\frac{m a_{\tau}^2D_{wn}}{2b_{\tau}\tau_{an}} \bigg]+\mathcal{O}(\tau_{an}^2,D_{wn}\tau_{an},\gamma\tau^3),  \label{AppCWKB1}  \\
    d_{\tau}&=\frac{\mu}{2D_{an}}\bigg(b_{\tau}-\frac{a_{\tau}D_{wn}}{2\tau_{an}} \bigg)  +\mathcal{O}(\tau_{an}^2,D_{wn}\tau_{an},\gamma\tau^3)\label{AppCWKB2}, 
\end{align}
and the small time-delay effect is implicitly contained in the auxiliary coefficients, 
\begin{align}
a_{\tau}&= \frac{\mu\big(1-\frac{\gamma\tau^2}{2m}\big)-\gamma\tau}{\mu\big(1-\frac{\gamma\tau^2}{2m}\big)^2} \label{ACoefa}, \\
    b_{\tau}&=a_{\tau}\bigg(1-\frac{\gamma\tau^2}{2m}\bigg). \label{ACoefb}
\end{align}
Note that $a_{\tau}$ and $b_{\tau}$ are coefficients bounded between 0 and 1, both approaching unity for a vanishing time delay. Following the singular perturbation techniques developed in Refs.~\cite{kosek19881,klosek19881,klosek19861,reimann19981}, we insert the ansatz into the FP equation and perform a perturbative expansion in the weak active-noise limit, which leads at leading order in $D_{an}$ to a Hamilton–Jacobi equation for $\Phi_{wkb}$. We further solve this equation by performing a Taylor expansion in $\tau_{an}$ and $D_{wn}$, which yields a hierarchy of differential equations from which we derive the effective quasipotential distribution in Eq. (\ref{QPDSD}) by employing the well-known method of characteristics. As a result we obtain $\Phi_{wkb}$ as a series in successive powers of the small parameters specified by conditions given in Eqs.~(\ref{PerWN_1}), (\ref{PerWN_2}), and (\ref{PerWN_3}). In App. \ref{App2}. we illustrate how this construction can be carried out step by step. For the case of a harmonic potential, Eq.~(\ref{QPDSD}) reduces to the exact solution of the harmonic potential.

The quasipotential distribution in Eq.~(\ref{QPDSD}) resembles that of a passive, non-delayed bistable Brownian particle subject to a (non-thermal) white noise, but with certain renormalized parameters. More concretely, we can obtain an estimation of its intrawell relaxation time (characteristic of the oscillatory motion within the stable potential well), that is
\begin{align}
    \omega_{ac}&=\sqrt{\frac{2 D_{an}}{m\mu}\bigg|\frac{\partial^2  \Phi_{wkb}}{\partial x^2}  \bigg|_{x=X_{1}^{\ast}}} \approx \sqrt{\frac{(\alpha-\gamma)}{m}\bigg(b_{\tau}-\frac{a_{\tau}D_{wn}}{2\tau_{an}} \bigg)}. \label{EqEIG_2}
\end{align}
By paying further attention, one may also realize that the time delay, the white-noise strength, and the active-noise correlation time enter through an effective mass $m_{\mathrm{eff}}$ and a modified potential barrier $\Delta U_{\mathrm{eff}}$, given by
\begin{align}
    m_{\mathrm{eff}}&=ma_{\tau}+\mu \tau_{an}b_{\tau}^2 -\frac{m a_{\tau}^2 D_{wn}}{2b_{\tau}\tau_{an}}, \label{EqPBEM} \\
    \Delta U_{\mathrm{eff}}&=\bigg(b_{\tau}-\frac{a_{\tau}D_{wn}}{2\tau_{an}} \bigg)\Delta U.
    \label{EqPBEU}
\end{align}
Interestingly, Eqs. (\ref{EqPBEM}) and ~(\ref{EqPBEU}) reveal that there is a competitive interplay among $\tau$, $\tau_{an}$ and $D_{wn}$ to govern the active particle escape dynamics. An increase in the white-noise strength lowers the effective barrier height for interwell transitions, so that larger values of $D_{wn}$ are expected to enhance the escape rate and to accelerate the hopping from one potential well to the other. A similar effect is produced by the time delay, which also reduces the effective barrier height and thus favors interwell transitions. This is precisely the trend observed in our analytical and numerical results presented in Secs. \ref{SubsectLRT} and \ref{SecNMS}. By contrast, increasing $\tau_{\mathrm{an}}$ makes the particle effectively heavier, thereby slowing down the escape process. 
Overall, the effect of the active-noise correlation time counteracts the effects of both white noise and time delay. Consequently, these observations provide a deeper insight into how active fluctuations can stabilize a state or modulate the energetic cost of jumping between potential wells, a feature that is essential for optimizing spectral amplification in SR regimes.

By using Eq. (\ref{AppWKBASS}), we can also compute the stationary mean square value of the spatial coordinate. The latter expresses up to leading order as follows, 
\begin{align}
     \langle x ^2\rangle_{st}&=
    \frac{\alpha-\gamma}{\beta}-\frac{D_{an} }{(\mu-\gamma\tau)(\alpha-\gamma)}\bigg( 2+\frac{D_{wn}}{\tau_{an}}\bigg)+\mathcal{O}(\tau_{an}^2,D_{wn}\tau_{an},\gamma\tau^2),
     \label{EqASXS}
\end{align}
which manifests that the spatial variance decreases for growing values of the time delay and the white-noise strength. 

Given the steady-state distribution function (\ref{QPDSD}), we can now evaluate approximately the noise-induced escape rate $k_{ac}$ by use of Kramer's procedure. Under the subsidiary conditions (\ref{PerWN_1}), (\ref{PerWN_2}) and (\ref{PerWN_3}); one obtains in the underdamped scenario (the interested reader can find the complete derivation in App. \ref{App4}),
\begin{align}
    k_{ac}&=  \frac{\lambda^{+}_0}{\sqrt{2}\pi}\Bigg[1-\frac{1}{2}\bigg( 1-\frac{\gamma\tau}{\mu}\bigg)\frac{\mu\tau_{an}}{m}\Bigg]e^{-\frac{(\mu-\gamma\tau)\Delta U }{2 D_{an}}\big(1-\frac{D_{wn}}{2\tau_{an}} \big)} +\mathcal{O}\bigg(\frac{(\gamma\tau)^2}{\mu^2},\frac{(\mu\tau_{an})^2}{m^2},\frac{D_{wn}^2}{\tau_{an}^2}\bigg),
    \label{MFRT_1}
\end{align}
where we have introduced the coefficient
\begin{align} 
\lambda^{\pm}_{0} = \frac{-\mu \pm \sqrt{\mu^{2}+4m(\alpha-\gamma)}}{2m}. 
\nonumber
\end{align} 
One may show that Eq. (\ref{MFRT_1}) returns the celebrated Kramers result for the escape rate \cite{kramers19401,hanggi19901,schuller20201,talkner19881,naeh19901} in the scenario of vanishing time delay, white noise, and active-noise correlation length when $D_{an}=\mu k_{b}T/2$, as expected. Equation (\ref{MFRT_1}) also retrieves at leading order in $\tau_{an}$ the results obtained previously in the scenario of bistable flow just driven by colored noise \cite{hanggi19851}.

Interestingly, the Arrhenius factor in (\ref{MFRT_1}) decreases for growing values of $D_{wn}/\tau_{an}$, which is consistent with previous observations: white noise and time delay tend to accelerate interwell hopping by lowering the effective activation barrier, whereas the active-noise correlation time tends to slow down the dynamics by increasing the effective inertia. Physically, as the memory of the active noise becomes more dominant relative to the (non-thermal) white noise, the system's escape dynamics are no longer governed by simple random hits but by a coherent temporal structure that reshapes the transition rates.  Additionally, since the Arrhenius factor decreases as $\tau$ increases, one may also realize that the escape rate is enhanced by time-delay effects, indicating that the SR of the system could be modulated by the lag in the nonlinear potential. This will be further discussed in Sec. \ref{SubsectLRT}. 

Additionally, Eq. (\ref{MFRT_1}) returns the escape rate in the overdamped scenario in agreement with the well-known result from rate-reaction theory \cite{hanggi19901,kramers19401} with $\tau=\tau_{an}=D_{wn}=0$ (see Eq. (\ref{MFRT_OD}) in App. \ref{App4}). We also note that this expression agrees qualitatively with previous overdamped results in the small colored-noise correlation-time limit \cite{reimann19981,klosek19881,fox19881,hanggi19841,woillez20201}, since it reproduces the same leading-order dependence on the correlation time. Nevertheless, the prefactor is not the same as somehow expected, since the present calculation is based on a different perturbative scheme: the active-noise correlation time is scaled differently, and the white-noise contribution is treated as a perturbation with respect to the active colored noise. Therefore, an exact correspondence with the overdamped small-correlation-time expressions should not be expected. For instance, although Ref.~\cite{doering19871} follows a procedure similar to ours, it adopts a constant colored-noise variance, while we address the situation where the integrated active-noise intensity is held constant, as specified in Eq.~(\ref{EqSN}). Instead, our result captures the dependence of the Arrhenius factor on the active-noise correlation time, in agreement with the well known colored-noise corrections to activated escape rates \cite{hanggi19841,fox19881,doering19871,klosek19881}.

Regarding the computation of the SPA, let us recall that we approximate the smallest non-vanishing eigenvalue of the FP equation by the escape rate \cite{hanggi19861,jung19882,anishchenko19991}, which yields
\begin{equation}
\lambda_{ac}\approx \lambda_{pass}\Bigg(1-\frac{\mu\tau_{an}}{2m}+\frac{\gamma\tau}{2m\lambda_{0}^{+}}\bigg(1-\frac{\mu}{2m\lambda_{0}^{+}+\mu} \bigg) \Bigg)e^{\frac{\mu \Delta U}{2D_{an}}\big( \frac{\gamma\tau}{\mu}+\frac{D_{wn}}{2\tau_{an}}\big)},
    \label{EqAER}
\end{equation}
where $\lambda_{pass}=\sqrt{2}\lambda_{0}^{+}/\pi e^{-\frac{\mu\Delta U}{2D_{an}}}$ coincides with the lowest non-vanishing eigenvalue retrieved by the classical Kramers escape rate for the scenario of a passive bistable particle just subjected to white noise \cite{hanggi19901}. Consistently, one may observe that the lowest non-vanishing eigenvalue gets exponentially increased by growing values of the time delay and the white-noise strength, thereby enhancing the interwell transition rate, in agreement with previous discussions. In contrast, increasing the active-noise correlation time reduces the lowest non-vanishing eigenvalue, but this reduction is less pronounced since it arises from an algebraic, rather than exponential, correction. In the following section, we demonstrate that this leads to a shift of the SR peak toward lower active-noise strengths when $\tau$ and $D_{wn}$ are increased. By contrast, increasing the active-noise correlation time $\tau_{an}$ shifts the peak toward higher active-noise strengths.

A comment is in order. To be valid the approximation (\ref{EqAER}) we focus the attention on the regime where the intrawell 
motion is much faster than the interwell transitions, such that the slow time-scale dynamics is expected to be dominated by rare noise-induced hopping between wells. Accordingly, this occurs for sufficiently small noise intensity \cite{hanggi19901,hanggi19931}, i.e., $\lambda_{ac}/\omega_{ac}\ll1$. By inspecting Eqs. (\ref{EqAER}) and (\ref{EqEIG_2}), one can see that this condition is guaranteed by the subsidiary conditions (\ref{PerWN_2}) and (\ref{PerWN_3}). For instance, this has been numerically observed in stochastic systems subject to white noise in overdamped \cite{matkowsky19811} as well as in underdamped \cite{kang20031} scenarios, or driven simultaneously by white and colored noises in the overdamped limit \cite{bartussek19951}. Hence, the conditions (\ref{PerWN_2}) and (\ref{PerWN_3}) guarantee the clear-cut separation needed for identifying the smallest non-vanishing eigenvalue with the escape rate provided by Eq. (\ref{MFRT_1}). In the following section we further discuss the validity of the linear response approximation (\ref{EqSuc}).  

\subsection{SPA in the linear response regime}\label{SubsectLRT}

\begin{figure}[h!]
\centering
\includegraphics[width=0.32\textwidth]{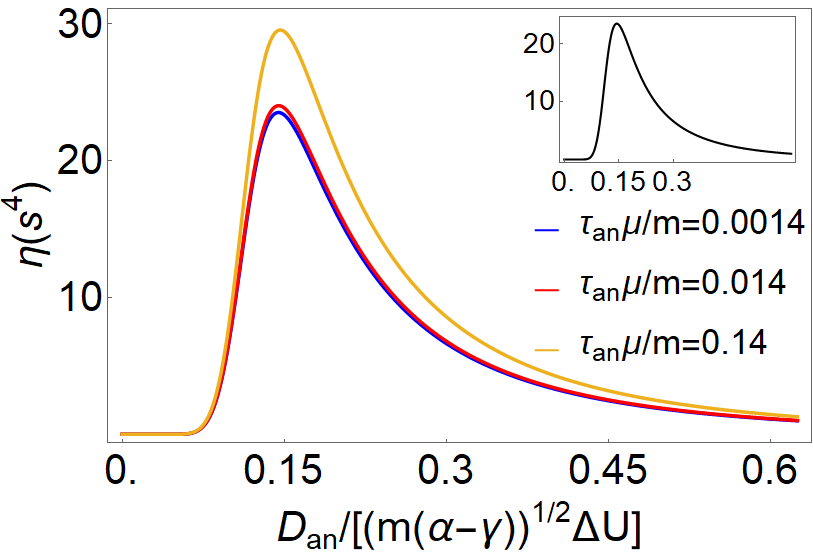} 
\includegraphics[width=0.32\textwidth]{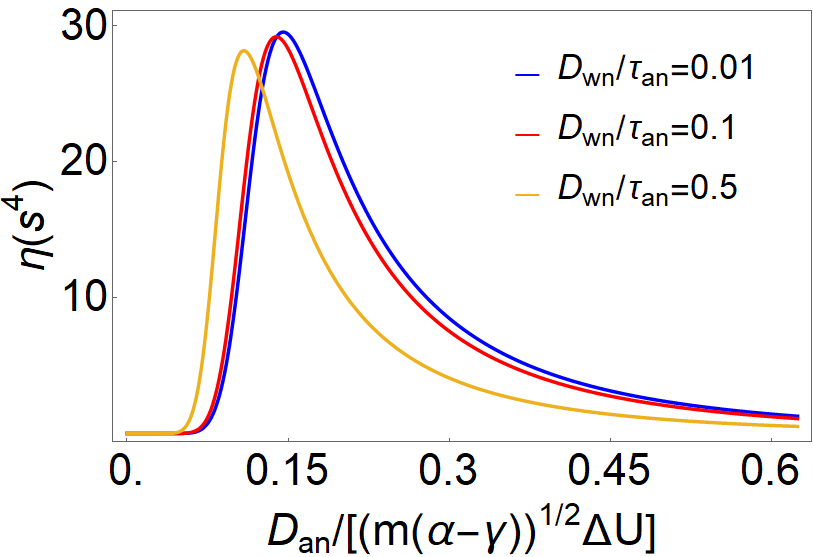} 
\includegraphics[width=0.32\textwidth]{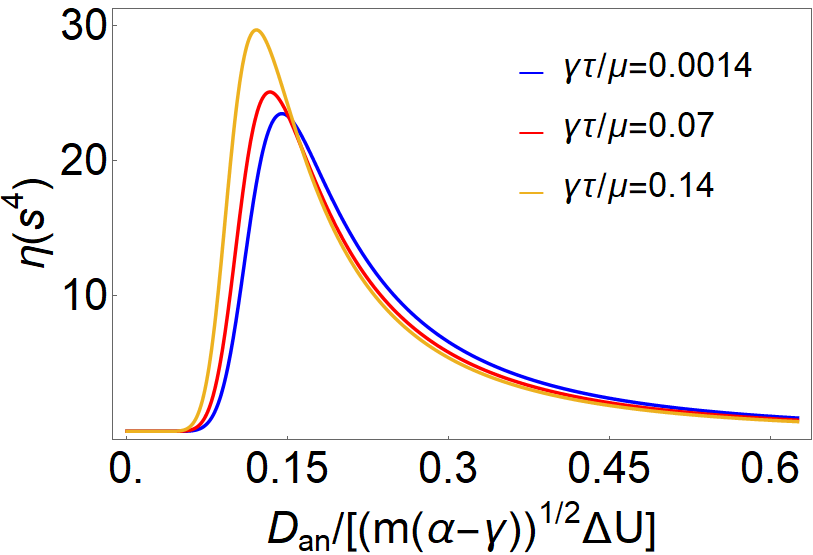} 
\includegraphics[width=0.32\textwidth]{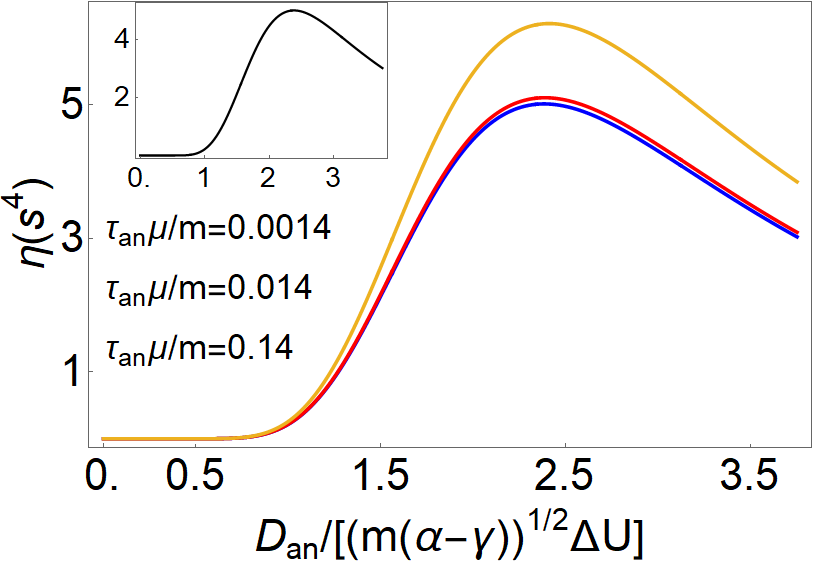} 
\includegraphics[width=0.32\textwidth]{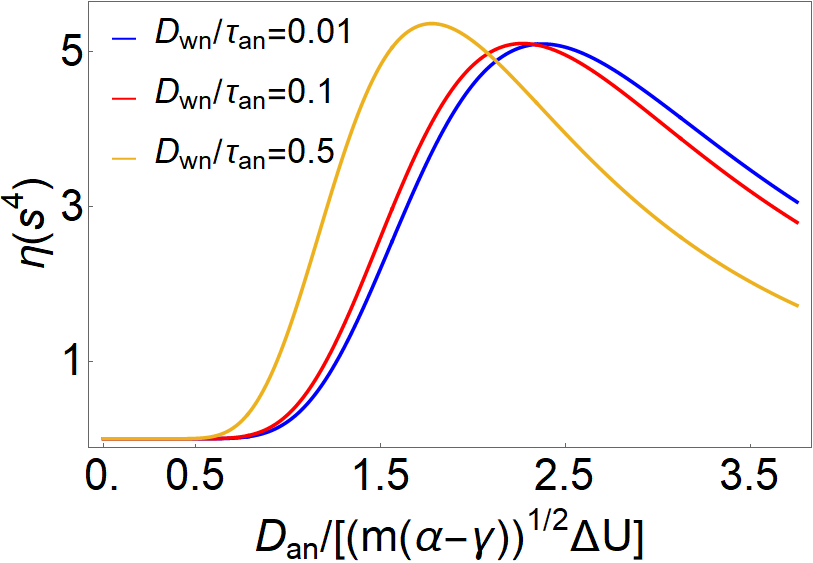} 
\includegraphics[width=0.32\textwidth]{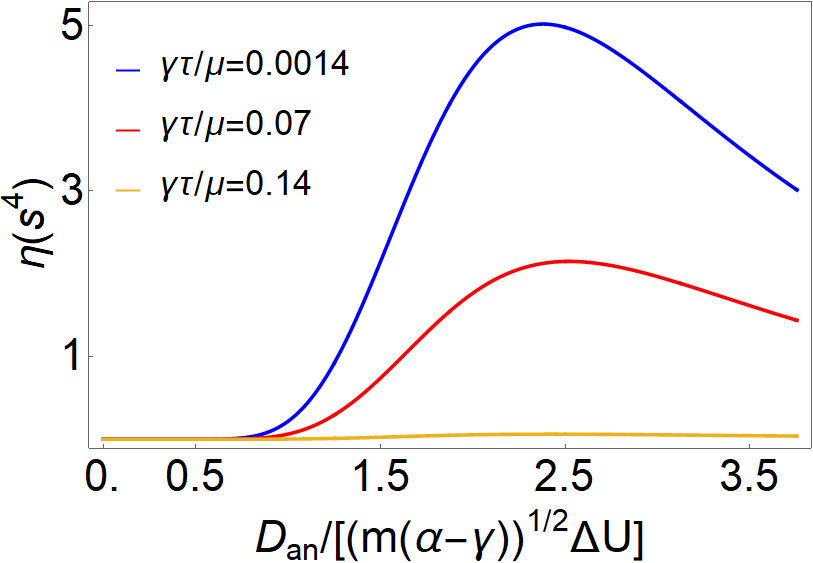} 
\caption{SPA as a function of the normalized active-noise strength $D_{an}$. The upper row corresponds to the \emph{underdamped} regime, $\mu\simeq\sqrt{m(\alpha-\gamma)}$, while the lower row shows the \emph{weakly overdamped} regime, $\mu\simeq10\sqrt{m(\alpha-\gamma)}$. Left panels: SPA for different values of the active-noise correlation time $\tau_{an}$, with time delay $\tau=0$ and white-noise strength $D_{wn}=0$. Center panels: SPA for different values of the white-noise strength $D_{wn}$, with fixed $\tau_{an}=0.14\,m/\mu$ (upper panel) and $\tau_{an}=0.014\,m/\mu$ (lower panel), and $\tau=0$. Right panels: SPA for different values of the time delay $\tau$, with $\tau_{an}=0$ and $D_{wn}=0$. The remaining parameters are fixed to $\Omega=0.01\,\mathrm{Hz}$, $\alpha=2.3\,\mathrm{kg\,s^{-2}}$, $\gamma=0.2\,\mathrm{kg\,s^{-2}}$, $\beta=0.01\,\mathrm{kg\,m^{-2}\,s^{-2}}$, and $m=1\,\mathrm{kg}$. The insets illustrate the results for the passive bistable system subjected to white noise, which has been included as a reference for comparison with the active bistable particle.}
\label{fig:EtaLR}
\end{figure}

Substituting the previous results into Eq.~(\ref{EqSuc}) yields an analytical expression for the SPA (the detailed derivation can be found in App. \ref{App4}). Let us remark that its validity requires that the frequency and strength of the driving force in comparison with
the potential well and the active noise strength are sufficiently small \cite{jung19911,gammaitoni19981}, that is
\begin{align}
    \frac{m^2F^2}{(\alpha-\gamma)\Delta U}\ll&1, \label{EqF1}\\
     \frac{m^{\frac{5}{2}}F^2}{(\alpha-\gamma)^{\frac{1}{2}}D_{an}}\ll&1,\label{EqF2} \\
     \frac{\Omega}{\omega_{ac}}\ll&1.  \label{EqF3}
\end{align}
Under these prescriptions, the SPA can be cast in the following form
\begin{align}
  \eta_{ac}(\tau,\tau_{an}, D_{wn},D_{an}) &\approx \eta_{pass}\bigg(1-2\frac{\gamma\tau}{\mu}\bigg)\Bigg(1+2 \frac{\mu\tau_{an}}{m}\bigg(1-\frac{\gamma\tau}{\mu} \bigg)-\frac{D_{wn}}{\tau_{an}} \bigg)\Bigg) \nonumber \\
  &\times \bigg[1-\frac{\Big(\Big(\frac{\pi\Omega}{\sqrt{2}\lambda^{+}_{0}}\Big)^2\frac{\mu\tau_{an}}{m}-\frac{\sqrt{2}\pi\Omega}{\lambda^{+}_{0}-2(\alpha-\gamma)\mu^{-1}}\frac{\gamma\tau}{\mu}\Big)e^{-\frac{\mu \Delta U}{D_{an}}\big(\frac{\gamma\tau}{\mu}+\frac{D_{wn}}{2\tau_{an}} \big)}}{1+\Big(\frac{\pi\Omega}{\sqrt{2}\lambda^{+}_{0}}\Big)^2e^{\frac{\mu \Delta U}{D_{an}}}} \bigg]
  \label{etaLRTapp}
\end{align}
where
\begin{equation}
    \eta_{pass}=\frac{m^2 \mu^2\Delta U}{4\beta D_{an}^2}\frac{1}{1+\Big(\frac{\pi\Omega}{\sqrt{2}\lambda^{+}_{0}}\Big)^2e^{\frac{\mu \Delta U}{D_{an}}}}, 
    \nonumber
\end{equation}
returns the SPA retrieved by the purely white-noise case that identifies with the passive bistable particle scenario. Interestingly, a first glance at the algebraic prefactor in Eq. (\ref{etaLRTapp}) reveals that the SR is enhanced by large active-noise correlation times in the underdamped scenario. This behavior is in contrast with previous results obtained for passive overdamped systems subject to colored noise \cite{hanggi19931,jia20011,luo20031,gammaitoni19892}. In particular, the influence of additive colored noise has also been investigated experimentally in sensory neurons. The authors of Ref. \cite{nozaki19991} showed in a FitzHugh-Nagumo neuronal model driven by colored noise that larger temporal noise correlations progressively weaken the resonant response. Paying attention to the susceptibility (\ref{EqSuc}) and the quasipotential coefficient (\ref{AppCWKB1}), such a discrepancy is attributed to the fact that a moderate increase of the active-noise correlation time may effectively grow the linear response of the bistable particle through its effect on the inertial dynamics. This can be intuitively understood from the interplay between the active-noise correlation time and the inertial time scale $m/\mu$: increasing $\tau_{an}$ shifts the active fluctuations toward lower frequencies, reducing the velocity variance and hence enhancing the coefficient $c_{\tau}$, thereby increasing the particle's linear response to the periodic forcing.

On the other hand, the white-noise contribution has a detrimental effect on the formation of SR, as it tends to reduce the coherence of the noise-assisted interwell transitions. The role of the time delay is more subtle, since $\tau$ affects both the mean square value of the spatial coordinate (\ref{EqASXS}) and the effective activation term appearing in the escape rate (\ref{MFRT_1}). As a consequence, the time delay can modify the optimal resonance condition not only through the exponential Arrhenius factor, but also through the algebraic prefactor that controls the transition rate. 

Figure~\ref{fig:EtaLR} illustrates SPA obtained within the linear-response regime as a function of the normalized active-noise strength in the underdamped (upper row) and weakly overdamped (lower row) scenarios. In all panels, $\eta$ displays a non-monotonic dependence on the active noise strength, which is the characteristic signature of SR \cite{gammaitoni19981}. Concretely, it reaches a pronounced maximum in the region $D_{\mathrm{an}}/\big(\sqrt{m(\alpha-\gamma)}\,\Delta U\big)\sim 0.14$--$0.18$. The left panel illustrates the effect of the active-noise correlation time $\tau_{\mathrm{an}}$. Increasing $\tau_{\mathrm{an}}$ enhances the maximum value of $\eta$, while leaving the position of the resonance peak almost unchanged. This indicates that increasing the active-noise correlation time enhances the SR response without significantly altering the optimal active-noise intensity. In contrast, increasing either the time delay $\tau$ or the white-noise strength $D_{\mathrm{wn}}$ shifts the resonance peak towards lower values of $D_{\mathrm{an}}$, suggesting that both parameters modify the matching condition responsible for stochastic resonance, which shall be further discussed below. It is also worth emphasizing that increasing the active-noise correlation time moderately enhances the SR response relative to the purely white-noise limit, as illustrated in the inner inset of the upper-left panel.

Paying attention to the effect of the time delay, this differs significantly between the underdamped and overdamped regimes. In the former, increasing $\tau$ produces only a modest enhancement of the SR peak. In contrast, in the overdamped regime, the resonant response gradually weakens as the time delay increases, eventually leading to the suppression of SR. This behavior coincides with the results of Du \emph{et al.}~\cite{du20111}, where increasing the delay weakens the synchronization between the noise-induced escape time and the deterministic forcing period, leading to a reduction of the resonance strength. By contrast, in our underdamped active system the time delay preserves the resonant mechanism while slightly enhancing the SR peak and shifting its optimum. This behavior is also distinct from that reported by previous works~\cite{gu20121}, where time-delayed feedback in an overdamped bistable system substantially increases the maximum signal-to-noise ratio without significantly changing the resonance position. A similarly nonuniversal influence of delay was reported for an overdamped bistable system driven by correlated additive and multiplicative noises, where the delay may either enhance or suppress the SNR and can even generate additional extrema depending on the noise intensities and cross-correlation strength~\cite{mei20091}. It is worth emphasizing that these studies consider different implementations of time-delayed feedback from the one employed here. In particular, our model incorporates a linear delayed restoring force in an underdamped active dynamics, whereas the delayed feedback mechanisms adopted in Refs.~\cite{du20111,gu20121,mei20091} are formulated within overdamped models. Consequently, the influence of the delay on SR is expected to arise from different physical mechanisms.

Lastly, we observe that increasing the white-noise strength lowers the height of the resonance peak associated with the active noise, confirming that white noise tends to degrade the amplification induced by active fluctuations, as revealed by Eq. (\ref{etaLRTapp}), and could eventually suppress the SR against $D_{an}$. 

\begin{figure}[h!]
\centering
\includegraphics[width=0.3\textwidth]{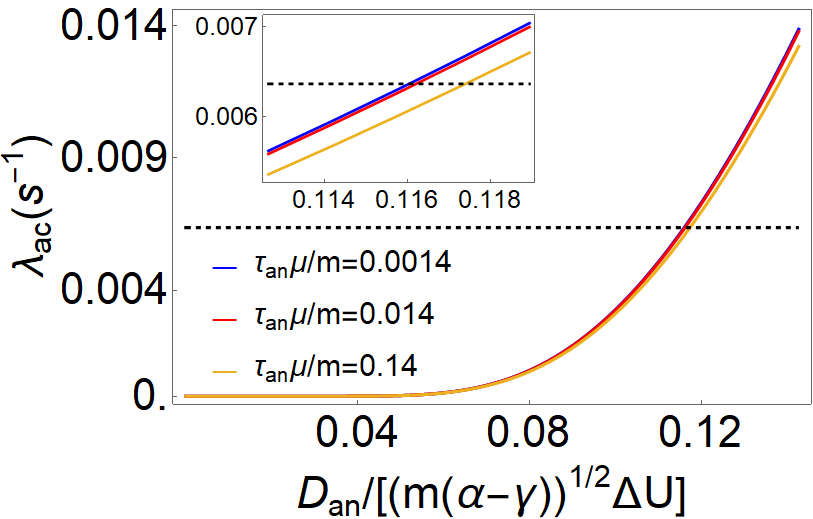} 
\includegraphics[width=0.3\textwidth]{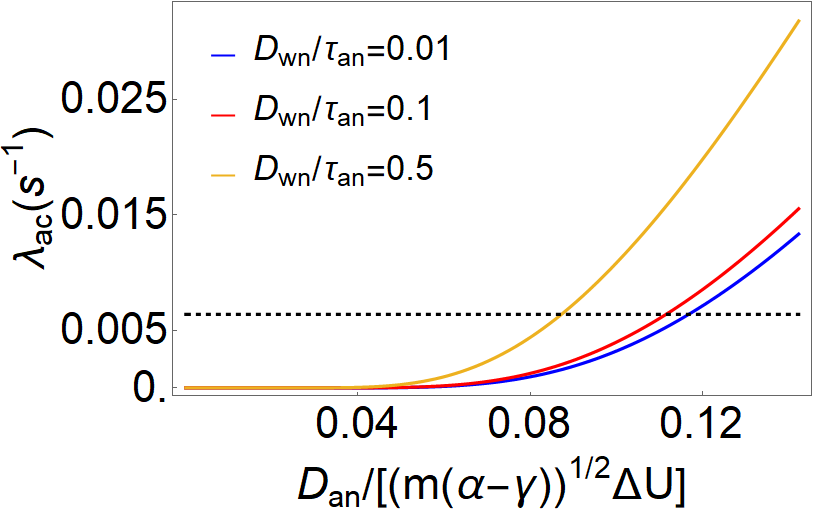} 
\includegraphics[width=0.3\textwidth]{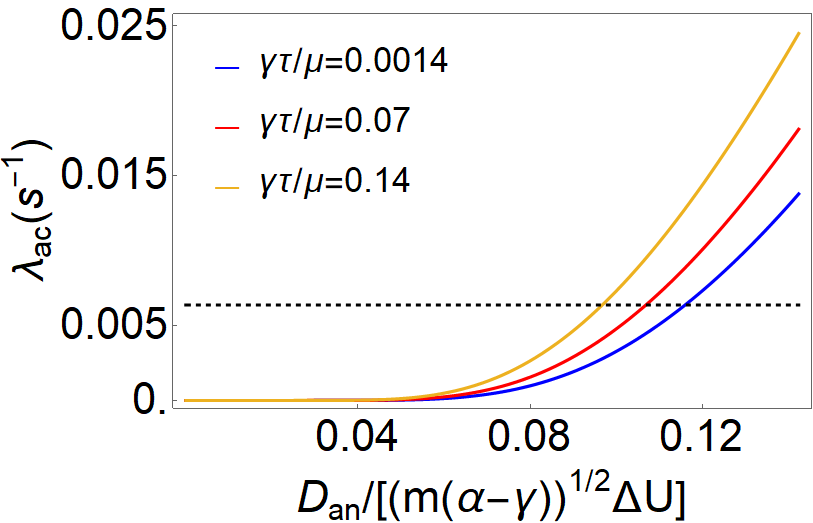} 
\caption{(Left) Lowest non-vanishing eigenvalue as a function of the normalized active-noise strength $D_{an}$ for different values of the active-noise correlation time with vanishing white-noise strength $D_{wn}=0$ and time delay $\tau=0$. The inset shows the SPA returned by the purely white-noise case, which identifies with the passive bistable particle scenario. (Central) Lowest non-vanishing eigenvalue as a function of $D_{an}$ for different values of the white-noise strength with fixed $\tau_{an}=0.14 m\mu^{-1}$ and $\tau=0$. (Right) Similarly, Lowest non-vanishing eigenvalue for different values of the time delay with $\tau_{an}=0$ and $D_{wn}=0$. We have fixed: $\Omega=0.01 \ Hz$, $\alpha=2.3 \ kg/s^2$, $\gamma=0.2 \ kg/s^2$, $\beta=0.01 \ kg/(ms)^2$, $\mu=1.4 \ kg/s$, $m=1 \ kg$.}
\label{fig:LNE}
\end{figure}

Additionally, one may be interested in the value of the SPA when the matching condition is satisfied, that is, $\Omega=(\pi/2) \lambda_{ac}$ \cite{gammaitoni19981}. According to Eq. (\ref{MFRT_1}), this is achieved for values of the active-noise strength, 
\begin{align}
        D_{an}&\approx D_{pass}\bigg(1- \frac{\gamma\tau}{\mu}-\frac{D_{wn}}{2\tau_{an}}\bigg)  \Bigg[1+\frac{\ln\big(1+\frac{\mu\tau_{an}}{2m}+\frac{\gamma\tau}{2m\lambda_{0}^{+}}\big(1-\frac{\mu}{2m\lambda_{0}^{+}+\mu}\big) \big)}{\ln\big(\frac{\sqrt{2}\Omega}{\lambda_{0}^{+}}\big)}     \Bigg]^{-1},
        \label{EqDOPT}
\end{align}
where $D_{pass}=\Delta U\mu/2\ln(\sqrt{2}\Omega/\lambda_{0}^{+})$ denotes the Kramer's estimate of the optimal noise strength in the purely white-noise case and in the absence of time delay, which identifies with the passive bistable particle scenario. As expected, Eq. (\ref{EqDOPT}) is consistent with the observations reported by Eq. (\ref{EqAER}) and in Fig.~\ref{fig:EtaLR}: both the time delay and the white-noise contribution shift the maximum of the SPA toward lower values of the active-noise intensity. Figure~\ref{fig:LNE} illustrates the lowest non-vanishing eigenvalue as a function of the active-noise strength for varying values of $\tau_{an}$, $\tau$, and $D_{wn}$. The crossing point with the dashed black horizontal line represents the matching condition, which is contained in the discussion about $\lambda_{ac}$ in Sec.~\ref{SubsectSSPDF}. In other words, the optimal value of $D_{\mathrm{an}}$ required to achieve SR decreases when either $\tau$ or $D_{\mathrm{wn}}$ is increased.

Finally, we obtain the value of the maximum retrieved by the SR condition, 
\begin{align}
   \eta_{max}(\tau,\tau_{an}, D_{wn},D_{an})\approx  \frac{4}{4+\pi^2}\frac{m^2 \mu^2\Delta U}{\beta (D_{an}^{pass})^2}\bigg(1-2\frac{\gamma\tau}{\mu}\bigg)\Bigg(1+2 \frac{\mu\tau_{an}}{m}\bigg(1-\frac{\gamma\tau}{\mu} \bigg)-\frac{D_{wn}}{\tau_{an}} \bigg)\Bigg), \label{EqSPA}
\end{align}
which clearly manifests that the white noise is harmful against the SR. Moreover, the enhancement produced by moderate active-noise correlation times is reduced as the delay time increases. This suggests that the delayed feedback partially counteracts the coherence induced by persistent active fluctuations, thereby weakening the increase of the SR peak associated with $\tau_{\mathrm{an}}$.

In summary, within the small time-delay regime, for small active-noise correlation times and sufficiently weak white-noise intensity, the SR mechanism persists in the presence of both white noise and time delay, although these agents may modify the matching condition. We find that the SR response may be slightly enhanced for moderate values of the active-noise correlation time. This enhancement can be intuitively understood as arising from the interplay between active persistence and the inertial time scale, which increases the particle's linear response to the periodic forcing. This behavior is in marked contrast with earlier findings for passive systems subject to colored noise in the overdamped limit , where increasing the colored-noise correlation time was found to reduce the height of the SR peak. In contrast, white noise reduces the amplitude of the SR peak and shifts it toward lower values of $D_{an}$, even though it promotes interwell transitions in the undriven case. The time delay has a more subtle effect: by modifying the effective escape dynamics, it shifts the value of the active-noise intensity at which the SR condition is fulfilled. In particular, both the time delay and the white-noise contribution displace the optimal active-noise intensity toward lower values, showing that they change the effective transition rate required to synchronize the bistable dynamics with the forcing frequency. Finally, let us emphasize that, in all the situations considered, the resonance peak systematically reaches larger values in the underdamped regime than in the overdamped one, suggesting that inertia favors the resonant response. This behavior can be intuitively understood in terms of the different dynamical roles played by inertia in the two regimes. In the overdamped limit, the particle rapidly relaxes toward the local potential minimum, whereas in the underdamped regime noise-induced fluctuations can cooperate with the particle's inertial motion to promote interwell transitions, thereby sustaining a stronger SR response.


\section{Numerical results}\label{SecNMS}

In this section, we present the numerical results obtained from direct simulations. We first consider the regime of small time delay and active-noise correlation time, where the perturbative theory developed in the previous section is expected to remain quantitatively accurate. The numerical simulations reproduce the non-monotonic SPA response and the location and parameter-dependent displacement of the resonant region predicted by the analytical theory. The agreement is particularly clear for the matching condition inferred from the lowest non-vanishing eigenvalue. By contrast, the absolute SPA amplitude is more sensitive to ensemble size and higher-order corrections, especially in its dependence on the active-noise correlation time. We also move beyond the perturbative regime by considering small-to-intermediate values of the time delay and active-noise correlation time, in order to explore the degrading effects upon the SR response due to stronger memory effects and environmental noise.

\subsection{Small time delay and active-noise correlation time}

\begin{figure}[h!]
\centering
\includegraphics[width=0.95\textwidth]{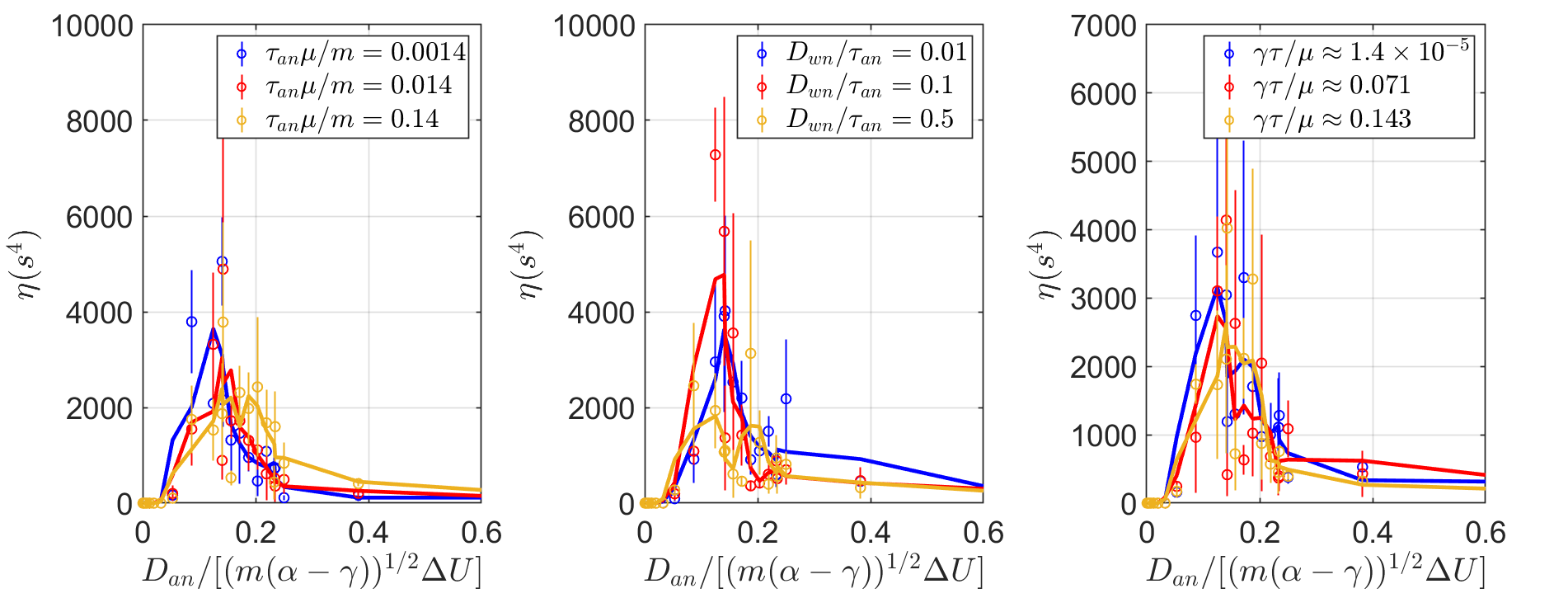}  
\caption{SPA as a function of the normalized active-noise strength.
The left panel illustrates the effect of the active-noise correlation time,
with $\tau_{\rm an}\mu/m=0.0014,\;0.014,\;0.14$,
$\tau=0$, and $D_{\rm wn}=0$.
The center panel shows the effect of the white-noise strength, with
$D_{\rm wn}/\tau_{\rm an}=0.01,\;0.1,\;0.5$,
$\tau_{\rm an}=0.1$, and $\tau=0$.
The right panel illustrates the effect of the time delay, with
$\gamma\tau/\mu\simeq1.4\times10^{-5},\;0.071,\;0.143$,
$\tau_{\rm an}=10^{-9}$, and $D_{\rm wn}=0$.
The remaining parameters are
$\Omega=0.01$ Hz, $\alpha=2.3$ kg\,s$^{-2}$,
$\gamma=0.2$ kg\,s$^{-2}$,
$\beta=0.01$ kg\,m$^{-2}$\,s$^{-2}$,
$\mu=1.4$ kg\,s$^{-1}$, $m=1$ kg, and $F=0.01$ N.
Error bars indicate the standard error of the mean (SEM) computed across
independent simulation blocks. The thicker curves represent a three-point
moving average of the same numerical data and are included solely as a visual
guide to highlight the resonant region.}
\label{fig:EtaNSTD}
\end{figure}

Figure~\ref{fig:EtaNSTD} presents the SPA as a function of the active-noise strength for different values of the active-noise correlation time, the time delay, and the white-noise strength. In all cases, the characteristic non-monotonic dependence associated with stochastic resonance is observed, with the resonance peak occurring at intermediate values of the normalized active-noise strength,
\begin{equation}
    \frac{D_{\mathrm{an}}}{\sqrt{m(\alpha-\gamma)}\,\Delta U}\approx0.14\text{--}0.18,
    \nonumber
\end{equation}
in good agreement with the analytical results shown in Fig.~\ref{fig:EtaLR}. This is further in quantitative agreement with the results of the matching condition manifested in Fig. \ref{fig:LNE}.

First, let us pay attention to the left panel which illustrates the influence of the active-noise correlation time. As predicted by the linear-response analysis, increasing $\tau_{\mathrm{an}}$ shifts the resonance region towards slightly larger values of the normalized active-noise strength and produces a broader resonance profile. Nonetheless, the statistical uncertainty prevents a clear identification of systematic changes in the peak amplitude that could confirm the analytical predictions.  

The right panel shows the effect of $\tau$. The characteristic non-monotonic feature of SR is preserved for all values of the time delay. Moreover, one may observe that increasing the time delay tends to enhance the resonance peak while also producing larger statistical fluctuations in its vicinity, in qualitative agreement with the theoretical predictions of the linear-response approximation.

Finally, the middle panel displays the influence of the white-noise strength. The addition of white noise leaves the location of the resonance maximum nearly unchanged, while slightly smoothing the resonance profile and reducing the peak amplitude. Consequently, sufficiently strong white noise weakens the SR response, again in agreement with the behavior predicted in the previous section.

It should be emphasized that the numerical estimates of the SPA amplitude are affected by finite-ensemble contributions, as discussed in Appendix \ref{App5}. Therefore, the numerical results provide stronger evidence for the existence and location of the stochastic-resonance region than for quantitative changes in its peak amplitude.

\begin{figure}[h!]
\centering
\includegraphics[width=0.95\textwidth]{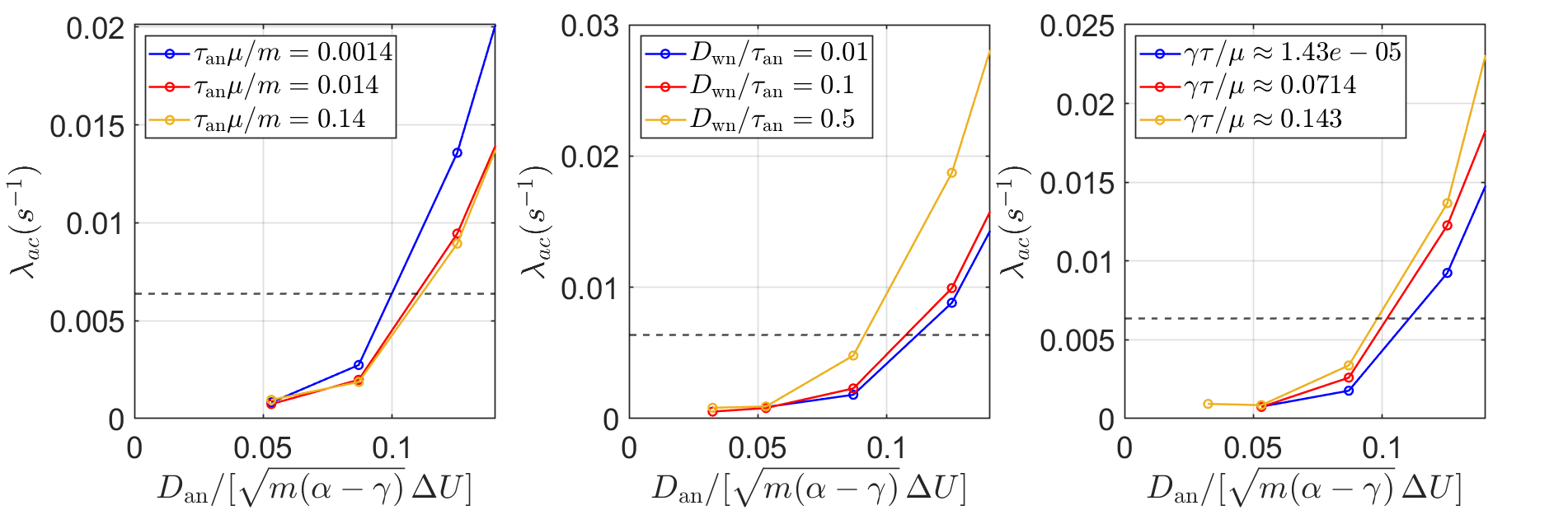} 
\caption{Lowest non-vanishing eigenvalue as a function of the normalized
active-noise strength. The left panel illustrates the effect of the
active-noise correlation time, with
$\tau_{\rm an}\mu/m=0.0014,\;0.014,\;0.14$,
$\tau=0$, and $D_{\rm wn}=0$.
The center panel shows the effect of the white-noise strength, with
$D_{\rm wn}/\tau_{\rm an}=0.01,\;0.1,\;0.5$,
$\tau_{\rm an}=0.1$, and $\tau=0$.
The right panel illustrates the effect of the time delay, with
$\gamma\tau/\mu\simeq1.4\times10^{-5},\;0.071,\;0.143$,
$\tau_{\rm an}=10^{-9}$, and $D_{\rm wn}=0$.
The remaining parameters are
$\Omega=0.01$ Hz, $\alpha=2.3$ kg\,s$^{-2}$,
$\gamma=0.2$ kg\,s$^{-2}$,
$\beta=0.01$ kg\,m$^{-2}$\,s$^{-2}$,
$\mu=1.4$ kg\,s$^{-1}$, and $m=1$ kg.
The horizontal dashed line indicates the stochastic-resonance matching
condition $\lambda_{ac}=2\Omega/\pi$.}
\label{fig:LNVEigenR}
\end{figure}

Additionally, Fig.~\ref{fig:LNVEigenR} presents the numerically estimated lowest non-vanishing eigenvalue for the same parameter families considered in Fig.~\ref{fig:LNE}, with the periodic forcing switched off for the MFPT calculation. A direct comparison between these figures provides numerical support for the analytical treatment of SR developed in the previous section. Remarkably, the analytical predictions show very good agreement with the numerical simulations, especially regarding the location and overall structure of the resonant region.

In all three cases, the escape rate, quantified by $\lambda_{ac}$, increases monotonically with the active-noise strength, in agreement with the analytical predictions shown in Fig.~\ref{fig:LNE}. The left panel shows that increasing the normalized active-noise correlation time, $\tau_{\mathrm{an}}$, systematically reduces the escape rate for a given $D_{\mathrm{an}}$. As a consequence, the matching condition is attained at larger active-noise intensities, consistently explaining the displacement of the SPA peak towards higher values of $D_{\mathrm{an}}$ predicted by the linear-response analysis and observed in the numerical simulations.

The right panel illustrates the effect of the delayed feedback. Increasing the normalized delay strength, $\gamma\tau/\mu$, enhances the escape rate, so that the matching condition is reached at lower values of the active-noise strength. This behavior is again fully consistent with the analytical results and accounts for the slight shift of the resonance peak towards smaller values of $D_{\mathrm{an}}$ observed in Figs.~\ref{fig:EtaLR} and~\ref{fig:EtaNSTD}.

Lastly, the middle panel shows that the three curves corresponding to different values of the normalized white-noise strength, $D_{\mathrm{wn}}/\tau_{\mathrm{an}}$, almost overlap, indicating that the additional non-thermal white noise has only a weak influence on the escape dynamics in the parameter regime considered. Nevertheless, increasing the white-noise intensity produces a slight increase in $\lambda_{ac}$, causing the matching condition to be satisfied at marginally lower active-noise strengths. This trend is consistent with the small displacement of the SPA maximum predicted analytically and confirmed by the numerical results.

\subsection{Small-to-intermediate regime}

\begin{figure}[h!]
\centering
\includegraphics[width=0.95\textwidth]{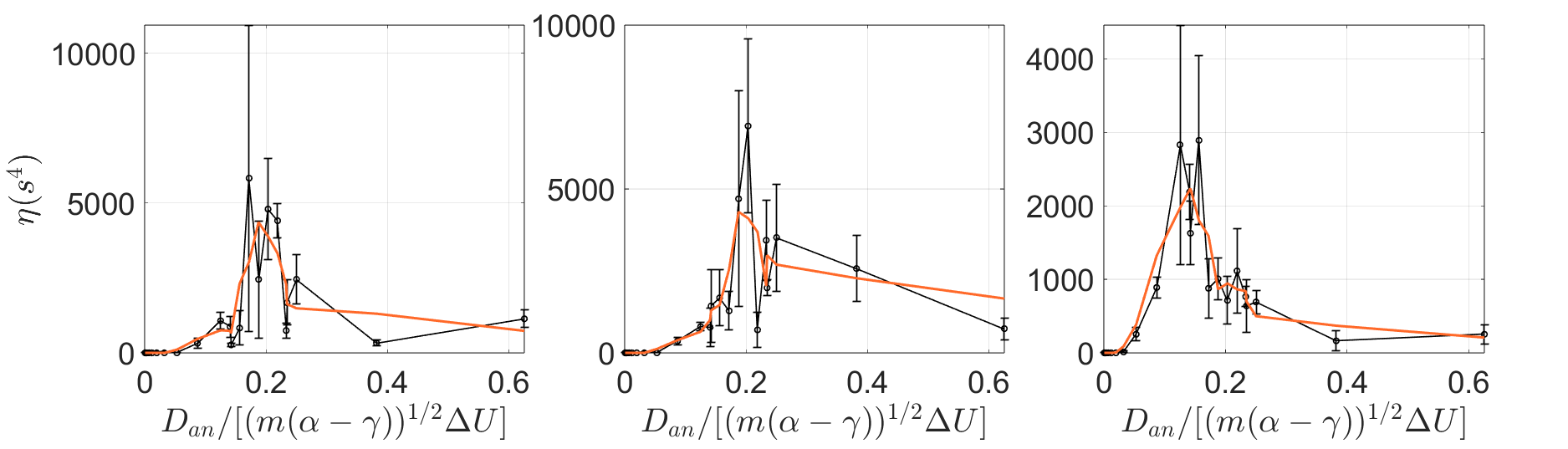} 
\caption{SPA as a function of the normalized active-noise strength. The left panel shows the effect of an intermediate active-noise correlation time, with $\tau_{an}\mu/m=1.4$, $\tau=0$, and $D_{\mathrm{wn}}=0$. The center panel illustrates the influence of the white-noise strength for $D_{\mathrm{wn}}/\tau_{an}=1$, with $\tau_{an}\mu/m=1.4$ and $\tau=0$. The right panel shows the effect of the time delay for $\tau\gamma/\mu=1.4$, with $\tau_{\mathrm{an}}\mu/m=1.4\times10^{-6}$ and $D_{\mathrm{wn}}=0$. The remaining parameters are fixed to $\Omega=0.01\,\mathrm{Hz}$, $\alpha=2.3\,\mathrm{kg\,s^{-2}}$, $\gamma=0.2\,\mathrm{kg\,s^{-2}}$, $\beta=0.01\,\mathrm{kg\,m^{-2}\,s^{-2}}$, $\mu=1.4\,\mathrm{kg\,s^{-1}}$, $m=1\,\mathrm{kg}$, and $F=0.01\,\mathrm{N}$. The orange line represents a three-point moving average of the same data, included solely as a visual guide to better highlight the position and shape of the resonant region in the presence of statistical fluctuations.}
\label{fig:STDSPA}
\end{figure}

Figure~\ref{fig:STDSPA} shows the SPA as a function of the normalized active-noise strength for intermediate values of the active-noise correlation time, the white-noise strength, and the time delay. Unlike the small-delay regime discussed in the previous section, where the resonance peak is sharp and well defined (see Fig. (\ref{fig:EtaLR})), all three cases exhibit a broader and less pronounced resonance profile. Increasing the active-noise correlation time, the white-noise strength, or the time delay progressively smooths the SPA curves and reduces their peak amplitude, while preserving the characteristic non-monotonic dependence on the active-noise strength. These results indicate that finite memory effects and stronger environmental noise gradually weaken the stochastic resonance response beyond the perturbative regime.

Nevertheless, the numerical simulations retain the main qualitative trends predicted by Eq.~(\ref{EqSPA}) in the perturbative scenario. In particular, it reveals that increasing either the time delay or the white-noise strength reduces the maximum SPA, in agreement with the numerical results. On the other hand, the broadening of the resonance profile observed for intermediate values of the active-noise correlation time reflects higher-order memory effects that are not captured within the perturbative approximation.

\section{Outlook and conclusions}\label{SecOC}

We have investigated SR in an AOUP evolving in a time-delayed bistable potential under periodic forcing and subject to an additional (non-thermal) white-noise source. By combining a perturbative analytical treatment based on linear-response theory with direct numerical simulations, we have characterized how the active-noise correlation time, the time delay, and the white-noise strength influence the synchronization between noise-induced barrier crossings and the external driving. The analytical predictions are further supported by direct numerical simulations within the perturbative regime.

Within the small-delay and small active-noise correlation-time regime, the analytical theory predicts that SR is preserved despite the simultaneous presence of white noise and delayed feedback. Interestingly, our analytical results suggest that the active-noise correlation time may slightly enhance the resonant response by increasing the SR peak, revealing that (persistent) active fluctuations promote synchronization between the interwell dynamics and the periodic forcing. In contrast, (non-thermal) white noise weakens the resonance and shifts the optimal active-noise intensity toward lower values, while the time delay produces a similar displacement of the resonance condition. 

Additionally, we find that time-delayed feedback has markedly different effects in the underdamped and overdamped regimes. In the latter, increasing the time delay progressively degrades the resonance response and may eventually suppress SR, whereas in the underdamped regime it can slightly enhance its response. Overall, these findings reveal that inertia qualitatively modifies the interplay between active persistence and time-delayed feedback in noise-activated dynamics. 

Extending the numerical analysis to intermediate values of the time delay and active-noise correlation time reveals that finite-memory effects progressively broaden and weaken the resonance peak. Thus, while the analytical theory captures the essential physics within the perturbative regime, higher-order memory effects become increasingly relevant beyond it, disrupting the synchronization between noise-induced interwell transitions and the periodic forcing and thereby deteriorating the SR response.

Several directions deserve further investigation. From a theoretical perspective, it would be interesting to extend the present perturbative approach to describe arbitrary delay times and active-noise correlation times, thereby capturing the full interplay between persistence and memory. On the experimental side, our predictions could be amenable to verification in active colloidal systems confined within engineered multistable energy landscapes. In particular, recent experiments on the escape dynamics of active particles in multistable potentials~\cite{militaru20211} demonstrate that noise-induced barrier crossing in active matter can already be accurately controlled and measured. Incorporating weak periodic forcing and delayed feedback into such experimental setups would provide a direct route to testing the stochastic-resonance mechanisms predicted here. More generally, the enhancement of SR induced by active persistence may provide useful guidelines for the design of synthetic active systems and microscopic devices in which transport, sensing, and spectral amplification can be optimized through the controlled interplay of activity, inertia, and memory.

\begin{acknowledgments}
This work was supported by the Spanish State Research Agency (AEI) and the European Regional Development Fund (ERDF, EU) under Project No. PID2023-148160NB-I00 (MCIN/AEI/10.13039/501100011033).
\end{acknowledgments}

\section*{Data Availability}
The data that support the findings of this study are available from the corresponding author upon reasonable request.

\appendix


\section{Linear stability analysis}\label{SecLSA}

In this section, we carry out a linear stability analysis of the nonlinear system of equations (\ref{EqDON0})-(\ref{EqCOU}). This system has two stable points $X^{*}_{1,2}$ and one saddle point $X^*_{0}$ with $\xi_{an}=0$ and $V=0$. By retaining first-order terms, after introducing the small perturbations around $X^*$ (i.e. $X(t) = X^* + \delta x(t),V(t) = \delta v(t), \xi_{an}(t) = \delta\xi_{an}(t) $ with $|\delta x| \ll |X^*|$, $|\delta v| \ll 1$ and 
$|\delta\xi| \ll 1$), we obtain the following linearized equations
\begin{align}
    \delta \dot{x} &= \delta v, \\[4pt]
m\,\delta \dot{v} 
&= -\mu\,\delta v 
+ \kappa\,\delta x(t) 
- \gamma\,\delta x(t-\tau) 
+ \sqrt{2D_{\mathrm{an}}}\,\delta \xi_{an}(t), \\[4pt]
\delta \dot{\xi}_{an} 
&= -\frac{1}{\tau_{\mathrm{an}}}\,\delta \xi_{an} .
\end{align}
where we have introduced the auxiliary coefficient evaluated at each fixed point,
\begin{equation}
    \kappa 
    = \begin{cases} 
        \alpha &  X_{0}^*, \\ 
         3\gamma-2\alpha &  X^*_{1,2}.
    \end{cases}
    \nonumber
\end{equation}
For the linear stability analysis, neglecting the transient noise forcing, we seek solutions of the form 
\(\delta x(t)= x_{0}e^{\lambda t}\) and 
\(\delta \xi_{an}(t)=\xi_{0} e^{\lambda t}\), 
which leads to the characteristic equation
\begin{align}
\bigg(\lambda+\frac{1}{\tau_{an}}\bigg)\Big[m\lambda^{2}+\mu\lambda-\kappa+\gamma e^{-\lambda\tau}\Big] &= 0.
\label{EqInst1}
\end{align}
Because $\xi_{an}$ enters the mechanical dynamics additively and is not fed back by $x$ or $v$ (i.e. $\delta \xi_{an}(t)$ is decoupled from the mechanical degrees of freedom), the augmented linearized system is separable. Hence, the active noise does not alter the deterministic stability boundary. As in the standard case, Eq. (\ref{EqInst1}) is a transcendental equation in $\lambda \in \mathbb{C}$, with infinitely many roots due to the time-delay term. 

In particular, in the vanishing time-delay scenario ($\tau=0$), the characteristic equation, Eq.~(\ref{EqInst1}), evaluated around the saddle point $X^*_{0}$ retrieves
\begin{align} 
\lambda^{\pm}_{0} = \frac{-\mu \pm \sqrt{\mu^{2}+4m(\alpha-\gamma)}}{2m}. 
\label{EqLP}
\end{align} 
This coincides with the result obtained in the Kramers analysis.
Clearly, if $\alpha>\gamma$, $\lambda^{+}_{0}$ is positive, and thus, $X_0^*$ remains unstable, which is consistent with the bistable structure. Similarly, we get for the potential wells
\begin{align}
\lambda=\frac{-\mu \pm \sqrt{\mu^{2}-8m(\alpha-\gamma)}}{2m},
\nonumber
\end{align}
so that the wells are stable (i.e., $\lambda>0$) provided that $\alpha>\gamma$ as well as $\mu>0$, which are fulfilled in the interesting scenario. 

Increasing the time delay $\tau$ may shift initially stable roots, $\operatorname{Re}(\lambda)<0$, toward the imaginary axis. Once these roots cross into the right half-plane, $\operatorname{Re}(\lambda)>0$, the potential wells lose their local stability, indicating that the delayed feedback can destabilize the bistable states \cite{cantisan20201}. On the other hand, the characteristic equation Eq.~(\ref{EqInst1})
evaluated at $\lambda = 0$ gives $-\kappa - \gamma = 0$, so the time delay cannot induce a real eigenvalue crossing. Consequently, the only 
delay-induced instability mechanism is a 
complex crossing $\lambda = \pm i\omega_c$ 
with $\omega_c \neq 0$, which constitutes a 
Hopf bifurcation. At the critical delay 
$\tau = \tau_c$, a pair of complex conjugate eigenvalues 
crosses the imaginary axis and a limit cycle of frequency 
$\omega_c$ emerges. From the physical point of view, the delay $\tau$ introduces a memory effect through the dependence of the potential at time $t$ on the position at $t-\tau$. This creates a feedback loop with a characteristic timescale $\tau$. When this feedback becomes strong 
enough, i.e., $\tau > \tau_c$, the system starts to 
oscillate at a frequency $\omega_c \sim 1/\tau$,  rather than simply converging, which would correspond to a real eigenvalue crossing.

Let us analyze this situation in further detail. By setting  $\lambda=i\omega_c$ in Eq.~(\ref{EqInst1}) and separating the real and imaginary parts, we obtain the following conditions,
\begin{align} 
\cos(\omega_c\tau) &= \frac{m\omega_c^{2}+\kappa}{\gamma}, \nonumber\\ 
\sin(\omega_c\tau) &= \frac{\mu\omega_c}{\gamma}, \label{Eqtauc} 
\end{align}
which can be compactly rewritten as 
\begin{align} 
\mu^{2}\omega_c^{2}+\left(\kappa+m\omega_c^{2}\right)^{2}
= \gamma^{2}.
\nonumber
\end{align} 
This expression constitutes a quartic equation for $\omega_c$ and admits a physical solution 
\begin{equation}
    \omega_c = \sqrt{ \frac{
    -\left(2m\kappa+\mu^2\right) 
    + \sqrt{\mu^4
    + 4m^2\gamma^2+4m\kappa\mu^2} }{2m^2}  },
    \label{eq:wc_positive}
\end{equation}
which is real and positive in the parameter range $\gamma^2>\kappa^2$. At the central fixed point $X_0^*$, this condition is incompatible with the instability property retrieved by Eq. (\ref{EqLP}). Therefore, the origin remains unstable for the parameter range considered, as required by the bistable structure underlying the SR mechanism.

In the case of the fixed points $X^{*}_{1,2}$, this condition is guaranteed when $\alpha/2<\gamma<\alpha<2\gamma$. This implies that, for $\tau > \tau_c$, the potential wells lose 
their stability via a Hopf bifurcation, and the system develops a deterministic limit 
cycle of frequency $\omega_c$. In this regime, the bistable dynamics is no longer sustained by the 
potential barrier, and consequently, the Kramers 
activation picture breaks down completely. The 
nonlinear dynamics is then dominated by the 
deterministic oscillation at $\omega_c$ rather than 
by noise-induced transitions between metastable 
states, rendering the standard SR framework 
infeasible. Furthermore, in the small-delay regime, Eq.~(\ref{Eqtauc}) gives $\tau_{c}=\mu/\gamma$. The stochastic resonance scenario, therefore, requires 
the fulfillment of the following conditions:
\begin{equation}
\begin{cases}
0<\tau < \frac{\mu}{\gamma}, & \text{if} \qquad \frac{\alpha}{2}<\gamma< \alpha
    \qquad \text{and}  \qquad  0<\mu,\\
0<\tau, & \text{if} \qquad 0<\gamma< \frac{\alpha}{2}
    \qquad \text{and}  \qquad  0<\mu,
    \label{eq:SR_conditions}
\end{cases}
\end{equation}
where the conditions on $\tau$ guarantee the existence of two stable fixed points $X^*_{1,2}$. Within this parameter window, noise-induced transitions between $X^*_{1}$ and $X^*_{2}$ remain the dominant dynamical mechanism, and the interplay between the periodic forcing, the white-noise intensity, the active-noise intensity, the active-noise correlation time, and the time delay can give rise to SR.

\section{Derivation of the time-delayed Fokker-Planck equation}\label{App1}

We introduce the active noise $\xi_{an}$ as an additional state variable, so that the extended process $(X,V,\xi_{an})$ is Markovian, whereas the reduced dynamics in $(X,V)$ remains non-Markovian for finite $\tau_{an}$. Instead, closed-form expressions for the kinetic equation in the reduced phase space spanned by $X$ and $V$ have been obtained using different approximation schemes. For instance, an approximate FP equation in the reduced phase space of $X$ and $V$ for the underdamped scenario was obtained by invoking a decoupling ansatz in \cite{hanggi19851} (i.e., $\tau_{an}\rightarrow 0$ and $\tau/D_{an}\ll 1$), whereas the overdamped situation has been addressed using the decoupling approximation \cite{hanggi19841}, the small-correlation-time expansion \cite{hanggi19941}, and the so-called unified-color-noise approximation \cite{hanggi19931,sharma20171} (i.e., $\tau_{an}\rightarrow 0$ or $\tau_{an}\rightarrow \infty$).

We start by obtaining the FP equation governing the time-evolution of the one-time probability density function $P_{1}(x,v,\xi,t)=\langle\delta(x-X(t))\delta(v-V(t))\delta(\xi-\xi_{an}(t))\rangle$ associated to the global system of variables $(X(t),V(t), \xi_{an}(t))$, that is 
\begin{align}
    \frac{\partial}{\partial t}P_{1}(x,v,\xi,t)&=-\bigg\langle\frac{\partial}{\partial x}\delta(x-X(t))\delta(v-V(t))\delta(\xi-\xi_{an}(t))\frac{V(t)}{m}\bigg\rangle-\bigg\langle\delta(x-X(t))\frac{\partial}{\partial v}\delta(v-V(t))\delta(\xi-\xi_{an}(t))\frac{d V(t)}{dt}\bigg\rangle \nonumber \\
    &-\bigg\langle\delta(x-X(t))\frac{\partial}{\partial v}\delta(v-V(t))\frac{\partial}{\partial \xi}\delta(\xi-\xi_{an}(t))\frac{d\xi_{an}(t)}{dt}\bigg\rangle \nonumber \\
    &=-v\frac{\partial}{\partial x} P_{1}(x,v,\xi,t) +\frac{1}{m}\frac{\partial}{\partial v}\bigg[\Big(\mu v+(\beta x^3-\alpha x-F\cos(\Omega t))-\sqrt{2D_{an}}\xi \Big)P_1(x,v,\xi,t) \bigg] \nonumber \\
    &-\frac{1}{\tau_{an}}\frac{\partial}{\partial \xi}\big[\big(\xi_{an}^{\infty}-\xi \big)P_{1}(x,v,\xi,t) \big] +\frac{\gamma}{m}\frac{\partial}{\partial v}\int\int  x_{\tau} P_{2}(x,v,\xi,t;x_{\tau},v_{\tau},t-\tau)d x_{\tau}dv_{\tau}  \nonumber \\
    &-\frac{\sqrt{2D_{wn}}}{m}\frac{\partial}{\partial v}\langle\delta(x-X(t))\delta(v-V(t))\delta(\xi-\xi_{an}(t))\Xi_{wn}(t)\rangle \nonumber \\
    &-\sqrt{2C_{an}'}\frac{\partial}{\partial \xi}\langle\delta(x-X(t))\delta(v-V(t))\delta(\xi-\xi_{an}(t))\Xi_{an}(t)\rangle.
    \label{AppFKEGI}
\end{align}
where $P_{2}(x,v,\xi,t;x_{\tau},v_{\tau},t-\tau)$ denotes the two-time probability density function, recall that $P_{2}(x,v,\xi,t;x_{\tau},v_{\tau},t-\tau)=\langle\delta(x-X(t))\delta(v-V(t))\delta(\xi-\xi_{an}(t))\delta(x_{\tau}-X(t-\tau))\delta(v_{\tau}-V(t-\tau))\rangle$. The last two terms in the right hand side of Eq. (\ref{AppFKEGI}) can be further simplified by using the Novikov's theorem applied to Gaussian noises \cite{hanggi19821}, i.e.
\begin{equation}
    \langle\xi_{a}(t)G[\xi_{a}]\rangle=\langle\xi_{a}(t)\rangle\langle G[\xi_{a}]\rangle+\int_{t_{0}}^{t}C_{a}(t-s)\bigg\langle\frac{\delta G[\xi_{a}]}{\delta \xi_{a}(s)}\bigg\rangle ds,
    \label{NVI}
\end{equation}
with $C_{a}(t-s)=\langle\xi_{a}(t)\xi_{a}(s)\rangle-\langle\xi_{a}(t)\rangle\langle\xi_{a}(s)\rangle$ for $t>s$. Replacing the mean value and covariance for the white noises $\Xi_{an}$ and $\Xi_{wn}$, characterized by Eqs. from (\ref{EQWN}) to (\ref{EQOUN}), yields the FP equation,
\begin{align}
    \frac{\partial}{\partial t}P_{1}(x,v,\xi,t)&=-v\frac{\partial}{\partial x} P_{1}(x,v,\xi,t) -\frac{1}{m}\frac{\partial}{\partial v}\bigg[\bigg(\sqrt{2D_{an}}\xi-\mu v-\Big(\beta x^3-\alpha x-F\cos(\Omega t)\Big)\bigg)P_{1}(x,v,\xi,t) \bigg] \nonumber \\
    &+\frac{C_{wn}'D_{wn}}{m^2}\frac{\partial^2}{\partial v^2} P_{1}(x,v,\xi,t)  -\frac{1}{\tau_{an}}\frac{\partial}{\partial \xi}\big[\big(\xi_{an}^{\infty}-\xi \big)P_{1}(x,v,\xi,t) \big] +\frac{C_{an}'}{\tau_{an}^2}\frac{\partial^2}{\partial \xi^2}P_{1}(x,v,\xi,t)\nonumber \\
    &+\frac{\gamma}{m}\frac{\partial}{\partial v}\int\int  x_{\tau} P_{2}(x,v,\xi,t;x_{\tau},v_{\tau},t-\tau)d x_{\tau}dv_{\tau}.
    \label{AppFKEG}
\end{align}
Here we have further employed the subsequent identities that follow from expressions (\ref{EqDON}) and (\ref{EqCOU}), i.e.
\begin{align}
     \frac{\delta X(t)}{\delta \Xi_{o}(s)}&=\Theta(t-s)\int_{s}^{t}\frac{\delta V(r)}{\delta \Xi_{o}(s)}dr,     \label{DF1}  \\ 
    \frac{\delta V(t)}{\delta \Xi_{o}(s)}&=\Theta(t-s)\bigg[\frac{\sqrt{2D_{wn}}}{m}\delta_{own}-\frac{\mu}{m}\int_{s}^{t}\frac{\delta V(r)}{\delta \Xi_{o}(s)}dr -\frac{1}{m}\int_{s}^{t}(-\alpha+3\beta x^2(r))\frac{\delta X(r)}{\delta \Xi_{o}(s)}dr +\frac{\sqrt{2D_{an}}}{m} \int_{s}^{t}\frac{\delta \xi_{an}(r)}{\delta \Xi_{o}(s)}dr\bigg] \nonumber\\
    &-\frac{\gamma}{m}\Theta(t-(s+\tau))\int_{s+\tau}^{t}\frac{\delta X(r-\tau)}{\delta \Xi_{o}(s)}dr, \label{DF2}\\
    \frac{\delta \xi_{an}(t)}{\delta \Xi_{o}(s)}&=\Theta(t-s)\delta_{o an}\bigg[\sqrt{2C_{an}'}-\frac{1}{\tau_{an}}\int_{s}^{t}\frac{\delta \xi_{an}(r)}{\delta \Xi_{o}(s)}dr\bigg],  \label{DF3} 
\end{align}
where $o=wn,an$ so that $\Xi_{o}$ represents the white noises. Notice that both noises are delta-correlated in time, after substituting in Novikov's identity (\ref{NVI}) the latter immediately entails: i) Eq. (\ref{DF1}) cancels, ii) whereas Eqs. (\ref{DF2}) and (\ref{DF3}) boil down to the first term in the right hand side. Replacing these results in (\ref{AppFKEGI}) yields the third and fifth terms in the right hand side of Eq. (\ref{AppFKEG}).

Clearly, from Eq. (\ref{AppFKEGI}) follows that the presence of a time delay makes the process intrinsically non-Markovian, so that the evolution 
of the one-time probability density depends on multi-time correlation 
functions: the kinetic description does not close at the 
level of the single-particle distribution but instead generates an infinite 
hierarchy of coupled equations \cite{frank20051}. In order to obtain a closed-form expression for the time-evolution of $P_{1}(x,v,\xi,t)$ (that is, a Markovian FP equation), we are enforced to approximate the last term in the right-hand side of Eq. (\ref{AppFKEG}). Here we focus the attention on the approximation scheme that holds in the small delay scenario, described in Sec. \ref{SecARSTD}. Concretely, whenever $\tau$ is sufficiently small compared with the dominant time scales, i.e. $\tau \Omega\ll 1$ and $\gamma\tau/\mu\ll 1$ are satisfied, the following Taylor approximation holds for the stochastic system under consideration \cite{guillouzic19991,frank20051},
\begin{equation}
    X(t-\tau)=X(t)-\tau V(t)+\frac{\tau^2}{2}\frac{d V(t')}{dt'}\Bigg|_{t'=t}+\mathcal{O}(\tau^3). \label{TESD}
\end{equation}
By replacing (\ref{TESD}) into the integral appearing in Eq. (\ref{AppFKEG}), after some manipulation we arrive to the following time-dependent FP equation

\begin{align}
    \frac{\partial}{\partial t}P_{1}(x,v,\xi,t)
    &=(\mathbb{L}_{FP}+\mathbb{L}_{ext}(t))P_{1}(x,v,\xi,t) \nonumber \\
    &=\big(\mathbb{L}_{0}+\mathbb{L}_{an}+\mathbb{L}_{td}
    +\mathbb{L}_{ext}(t)\big)P_{1}(x,v,\xi,t),
    \label{FPEq}
\end{align}
that completely contains the small-time delay effects. We have identified the standard unperturbed FP and the periodic-driving operators, which deduce directly from the first and second lines in the right hand side of Eq. (\ref{AppFKEG}), respectively,
\begin{align}
    \mathbb{L}_{0}
    &= -v\frac{\partial}{\partial x}
    +\frac{\partial}{\partial v}
    \bigg(\frac{\mu}{m} v+\frac{1}{m}\frac{\partial U_{0}(x,t)}{\partial x}\bigg)
    +\frac{C_{wn}'D_{wn}}{m^2}\frac{\partial^2}{\partial v^2},
    \label{DOFP1} \\
    \mathbb{L}_{ext}(t)
    &= -\frac{F}{m}\cos(\Omega t)\frac{\partial}{\partial v},
\end{align}
together with the differential operator that governs the active-noise dynamics,

\begin{align}
 \mathbb{L}_{an}
 &= -\frac{\sqrt{2D_{an}}}{m}\frac{\partial}{\partial v}\xi
 -\frac{1}{\tau_{an}}\frac{\partial}{\partial \xi}
 \bigg[\big(\xi_{an}^{\infty}-\xi \big)
 -\frac{C_{an}'}{\tau_{an}}\frac{\partial}{\partial \xi}\bigg],
 \label{DOFP2}
\end{align}
whereas the last term in Eq.~(\ref{FPEq}) contains the effects of the time delay,

\begin{align}
    \mathbb{L}_{td}
    &=-\frac{\gamma}{m}\frac{\partial}{\partial v}
    \bigg[
    \tau v+\frac{\tau^2}{2m}
    \bigg(
    \mu v+\frac{\partial U_{0}(x,t)}{\partial x}
    +\frac{C_{wn}'D_{wn}}{m}\frac{\partial}{\partial v}
    -\sqrt{2D_{an}}\xi
    \bigg)
    \bigg],
    \label{DOFPSD}
\end{align}
Here, $U_{0}$ denotes the nonlinear potential in Eq.~(\ref{PtFunc}) for a vanishing time delay. Note that Eq.~(\ref{FPEq}) reduces to the bistable flow driven by colored noise when $\tau$, $F$, and $D_{wn}$ vanish.

\section{WKB derivation of the steady-state probability distribution}\label{App2}
In this section, we illustrate the procedure to obtain the approximated expression (\ref{QPDSD}) for the steady-state probability density function. Essentially, we shall restrict our study to the parameter set where a WKB perturbative procedure provided in Ref.~\cite{kosek19881} can return a reliable asymptotic solution. Let us first pay attention to the linear scenario in which the potential is quadratic and there is no external driving, i.e. $\beta=0$ and $F=0$; and we restrict ourselves to the scenario where the small delay approximation is sufficient at first order. Hence, under appropriate conditions, the system described by Eq.~(\ref{FPEq}) will reach a steady state in the asymptotic time whose probability density function, say $P_{st}$, can be exactly computed, that is
\begin{equation}
   P_{st}(x,v,\xi)=\frac{(4\pi)^3}{\sqrt{\det\Phi}}\text{exp}\big(-\vect x^T\vect \Phi^{-1}\vect x\big),
   \label{EqHPDS}
\end{equation}
where $\vect x=(x,v,\xi)^{T}$ (with $T$ denoting the usual transpose operation) and we have introduced the potential function matrix $\vect \Phi$, whose inverse is given by
\begin{equation}
    \vect \Phi=\left(\begin{array}{ccc}
      \frac{2}{\mu\alpha}\big[C_{wn}'D_{wn}+\frac{2C_{an}'D_{an}(m+\mu\tau_{an})}{m+\tau_{an}(\mu+\tau_{an}\alpha)} \big]   & 0 & \frac{2\tau_{an} C_{an}'\sqrt{2 D_{an}}}{m+\tau_{an}(\mu+\alpha\tau_{an})}\\
      0  & \frac{2}{\mu}\big[\frac{C_{wn}'D_{wn}}{m}+\frac{2C_{an}'D_{an}}{m+\tau_{an}(\mu+\tau_{an}\alpha)} \big]   & \frac{2 C_{an}'\sqrt{2 D_{an}}}{m+\tau_{an}(\mu+\alpha\tau_{an})}\\
        \frac{2 \tau_{an} C_{an}'\sqrt{2 D_{an}}}{m+\tau_{an}(\mu+\alpha\tau_{an})}  &\frac{2 C_{an}'\sqrt{2D_{an}}}{m+\tau_{an}(\mu+\alpha\tau_{an})} & \frac{2C_{an}'}{\tau_{an}}
    \end{array}\right),
    \label{InQPFH}
\end{equation}
where we have obtained $\vect \Phi$ after solving the Lyapunov equation $\vect \Gamma\vect \Phi-\vect \Phi\vect \Gamma=4\vect D$, once we have introduced the dissipative $\vect \Gamma$ and diffusion $\vect D$ matrices.

Now we turn to the derivation of an approximate solution for the steady-state of Eq.~(\ref{FPEq}) which holds up to leading order in active noise parameters and for sufficiently weak white noise. Before we proceed further, it is convenient to perform the following change of variables 
\begin{align}
    v\rightarrow\bigg(\frac{\Delta U}{m}\bigg)^{\frac{1}{2}}v, \ \ x\rightarrow\bigg(\frac{\Delta U}{\alpha-\gamma}\bigg)^{\frac{1}{2}}x, \ \ t\rightarrow\bigg(\frac{m}{\alpha-\gamma}\bigg)^{\frac{1}{2}}t, \ \ U_{0}\rightarrow\Delta U U_{0},
\end{align}
where we can discern the range of validity of our approximations; concretely, we get a Kramer equation with the following linear differential operators 
\begin{align}
   \mathbb{L}_{0}&= -v\frac{\partial}{\partial x}+\frac{\partial}{\partial v}\bigg(\bigg(\frac{\mu^2}{m(\alpha-\gamma)}\bigg)^{\frac{1}{2}} v+m\frac{\partial U_{0}(x,t)}{\partial x}\bigg)+\bigg(\frac{C_{wn}'^2D_{wn}^2}{m(\alpha-\gamma)\Delta U^2}\bigg)^{\frac{1}{2}}\frac{\partial^2}{\partial v^2}, \nonumber \\
    \mathbb{L}_{an}&= -\bigg(\frac{2D_{an}}{(m (\alpha-\gamma))^{\frac{1}{2}}\Delta U}\bigg)^{\frac{1}{2}}\frac{\partial}{\partial v}\xi+\bigg(\frac{m}{\tau_{an}^2(\alpha-\gamma)}\bigg)^{\frac{1}{2}}\frac{\partial}{\partial \xi}\bigg[\xi +\frac{m^{\frac{1}{4}}}{\tau_{an}(\alpha-\gamma)^{\frac{1}{4}}}\frac{\partial}{\partial \xi}\bigg], \label{APPDOAN} \\
       \mathbb{L}_{td}^{\text{SD}}&=-\bigg(\frac{\gamma^2\tau^2}{m(\alpha-\gamma)}\bigg)^{\frac{1}{2}}\frac{\partial}{\partial v}\bigg[ v+\frac{\tau^2}{2m}\bigg(\mu v+\big(m(\alpha-\gamma)\big)^{\frac{1}{2}}\frac{\partial U_{0}(x,t)}{\partial x}+\frac{C_{wn}'D_{wn}}{\Delta U}\frac{\partial}{\partial v} \nonumber \\
       &-\bigg(\frac{2(m(\alpha-\gamma))^{\frac{1}{2}}D_{an})}{\Delta U}\bigg)^{\frac{1}{2}}\xi\bigg)\bigg],  \label{APPDOTD} 
\end{align}
within these change of variables the external force switches as $F\rightarrow\big(\frac{m^2F^2}{(\alpha-\gamma)\Delta U}\big)^{\frac{1}{2}}$. Paying attention to Eq. (\ref{APPDOAN}) we realize the perturbative parameter $\epsilon$ in the weak active-noise regime, that is 
\begin{equation}
    \epsilon=\frac{D_{an}}{(m (\alpha-\gamma))^{\frac{1}{2}}\Delta U} \ll 1,
    \nonumber
\end{equation}
Furthermore, in order to make a fair comparison between the effects arising from active and white noises (we impose the noise strength to be comparable in Eq.~(\ref{EqDON})) and Eqs.~(\ref{EQWN}) and (\ref{EQOUN}) suggest choosing $C_{wn}'=\frac{D_{an}}{\tau_{an}}$. Physically, this choice guarantees that the active-noise effects are not overshadowed by the white noise effects, as stated in Sec.~\ref{SecMMT},  which returns the following perturbative parameter for the weak white-noise regime
\begin{equation}
    \delta =\frac{D_{an}D_{wn}}{(m(\alpha-\gamma))^{\frac{1}{2}}\Delta U\tau_{an}}=\frac{\epsilon D_{wn}}{\tau_{an}}\ll 1,
    \nonumber
\end{equation}
which is further satisfied by setting
\begin{equation}
   \frac{D_{wn}}{\tau_{an}}\ll 1.
   \nonumber
\end{equation}
Similarly, we also realize that the influence of the correlation time scale of the active noise in (\ref{APPDOAN}) also constitutes a perturbative parameter in the semi-Markovian regime, as expected
\begin{equation}
    \nu=\frac{\tau_{an}(\alpha-\gamma)^{\frac{1}{2}}}{m^{\frac{1}{2}}}\leq\frac{\tau_{an}\mu}{m}\ll 1    
\end{equation}
Recall that the strength of the time-delay effects in Eq. (\ref{TESD}) also represents a perturbative parameter, see Eq. (\ref{PerTD}).

Here we illustrate the derivation of the approximate solution (\ref{QPDSD}) by following a perturbative approach in the perturbative parameter $\epsilon$ in the time domain of interest, i.e. $\epsilon\ll 1$.  More specifically, we assume that the strength of the active noise is sufficiently small compared to the height of the potential barrier. For seek of clarity, it is convenient to make the following renormalization of the differential operator coefficients 
\begin{equation}
    \mu\rightarrow\mu (m(\alpha-\gamma))^{\frac{1}{2}},
    D_{an}\rightarrow\frac{\epsilon \mu}{2},
\end{equation}
and the change of variable $\xi\rightarrow \frac{m}{\sqrt{2D_{an}}} \xi$. Hence, we can approximate the steady-state solution by the well-known WKB method, which consists of seeking solutions to Eq.(\ref{FPEq}) in the so-called WKB form \cite{naeh19901,kosek19881}, i.e.
\begin{equation}
     P_{st}(x,v,\xi)\approx K(x,v,\xi)e^{-\frac{\phi(x,v,\xi)}{\epsilon}},
    \label{AppWKBA}
\end{equation}
where $\phi(x,v,\xi)$ plays the role of a quasipotential function and $K(x,v,\xi)$ is a boundary layer function. Both functions are obtained by firstly inserting the WKB ansatz (\ref{AppWKBA}) into the Fokker-Planck equation in the small time-delay regime (\ref{FPEq}), and then, performing a Taylor expansion in terms of $\epsilon$. More specifically, by collecting the $\mathcal{O}(1)$ terms we get a Hamilton-Jacobi (eikonal) type equation rather difficult to solve, while $K(x,v,\xi)$ satisfies an homogeneous transport equation \cite{naeh19901}. Instead we can further proceed by expanding the quasi-potential function in terms of the white noise strength:
\begin{equation}
    \phi(x,v,\xi)=\phi_{0}(x,v,\xi)+\phi_{1}(x,v,\xi)\bigg(\frac{D_{wn}}{m}\bigg)+\phi_{2}(x,v,\xi)\bigg(\frac{D_{wn}}{m}\bigg)^2+\mathcal{O}(\delta^3).
    \label{AppWKA1}
\end{equation}
Similarly as before, we replaced (\ref{AppWKA1}) into the aforementioned Hamilton-Jacobi equation and set the coefficients of each power of $D_{wn}$ separately to zero, returning a recursive system of equations for $\phi_{i}$ (with $i=1,\cdots,\infty$) \cite{kosek19881}. The leading order retrieves 
\begin{align}
\frac{2\mu}{\tau_{an}^2}\phi_{0\xi}^2-\frac{2\xi}{\tau_{an}}\phi_{0\xi}+2v\phi_{0x}+f_{\tau}(x,v,\xi)\phi_{0v}=0,
     \label{AppWKA2}
\end{align}
where we have introduced the auxiliary function $f_{\tau}(x,v,\xi)=2(\xi-U_{x}-\mu v)+2\gamma \tau v+\gamma\tau^2/m(\mu v+U_{x}-\xi)$ and used the short notation $\phi_{ia}=\frac{\partial\phi_{i}}{\partial a}$ with $a=\xi,x,v$ as well as $U_{x}=\frac{\partial U_{0}}{\partial x}$. Equation (\ref{AppWKA2}) is a first-order non-linear differential equation of the  Ricatti type and coincides with the result obtained in \cite{kosek19881} when we neglect the time delay and the asymptotic active noise effects (i.e. $\tau\rightarrow 0$ and $\xi_{an}^{\infty}\rightarrow 0$). Now we follow the prescription proposed in Ref. \cite{kosek19881} to obtain an approximated solution of Eq. (\ref{AppWKA3}) up to second order in the correlation time $\tau_{an}$ of the active noise (i.e. $\mathcal{O}(\nu)$) in the same fashion as before. Basically, this consist on replacing the following ansatz in Eq. (\ref{AppWKA2}),
\begin{equation}
    \phi_{0}(x,v,\xi)=\psi_{0}(x,v,\xi)+\psi_{1}(x,v,\xi)\Big(\frac{\tau_{an}}{m}\Big)+\psi_{2}(x,v,\xi)\Big(\frac{\tau_{an}}{m}\Big)^2+\mathcal{O}(\nu^3),
    \label{AppWKA2a}
\end{equation}
and collecting terms $\mathcal{O}(\nu^{n})$ for $n=0,1,2$ to get recursive relations for $\psi_{n}$ that are easier to solve. For instance, the leading order returns $\psi_{0\xi}=0$, and thus, $\psi_{0}$ is independent of the active noise variable (which coincides with the solution found in \cite{kosek19881}); whereas the second-order is identically satisfied. The next leading order yields,
\begin{align}
    -2\mu \psi^{2}_{1\xi}-f_{\tau}(x,v,\xi)\psi_{0v} +2\xi\psi_{1\xi}-2v\psi_{0x}=0.
    \label{AppWKA3}
\end{align}
Now we demmand that $\psi$ is a real function in $\xi$ so that the discriminant of the quadratic equation (\ref{AppWKA3}) for $\psi_{1\xi}$ must be nonnegative for any value of $x$ and $v$, we find
\begin{align}
    \psi_{0}&=a_{\tau}\frac{v^2}{2}+b_{\tau}U_{0}(x), \nonumber \\
    \psi_{1}&=\frac{\xi^2}{2\mu}-b_{\tau}\xi v+r_{1}(x,v),
    \label{AppWKA4}
\end{align}
with the auxiliary coefficients defined in Eqs. (\ref{ACoefa}) and (\ref{ACoefb}), and where $r_{1}(x,v)$ is an arbitrary function that must be determined. The latter can be done by replacing the solution (\ref{AppWKA4}) in the expression retrieved by the next leading order in $\tau_{an}$, i.e.
\begin{align}
2(\xi-2\mu\psi_{1\xi})\psi_{2\xi}-f_{\tau}(x,v,\xi)\psi_{1v}-2v\psi_{1x}=0.
   \label{AppWKA5}
\end{align}
From here we can exactly integrate for $\psi_{2}$, i.e.
\begin{equation}
    \psi_{2}=\int\frac{f_{\tau}(x,v,\xi)\psi_{1v}+2v\psi_{1x}}{2(\xi-2\mu\psi_{1\xi})}d\xi+r_{2}(x,v),
    \label{AppWKA6}
\end{equation}
where $r_{2}(x,v)$ is another arbitrary function to be determined by applying the next order. By paying attention to (\ref{AppWKA6}), we must demand that $\psi_2$ is nonsingular for the value $\xi_{0}=2\mu b_{\tau}v$ when $\xi_{0}-2\mu\psi_{1\xi}(x,v,\xi_{0})=0$, which yields a partial differential equation for $r_{1}$:
\begin{equation}
    f_{\tau}(x,v,\xi_{0})\big(r_{1v}-b_{\tau}\xi_{0}\big)+2vr_{1x}=0,
    \label{AppWKA7}
\end{equation}
where $f_{\tau}(x,v,\xi_{0})=(\gamma\tau^2/m-2)U_{x}+\frac{\mu a_{\tau}}{2}(\gamma\tau^2/m-2)^2v$.
After replacing the result for $\xi_{0}$, we can solve the above differential equation by using the characteristic method; that is, Eq. (\ref{AppWKA7}) can be recast as follows
\begin{equation}
\frac{dr_{1}}{dt}=2\mu b_{\tau}^2 v\frac{dv}{dt},
    \label{AppWKA8}
\end{equation}
where $\frac{dx}{dt}=v$ and $\frac{dv}{dt}=f_{\tau}(x,v,\xi_{0})/2$. Hence the solution of (\ref{AppWKA8}) is given by \cite{kosek19881}
\begin{equation}
    r_{1}=\mu b_{\tau}^2v^2, 
    \nonumber
\end{equation}
which must be substituted in (\ref{AppWKA4}), i.e.
\begin{equation}
    \psi_{1}=\frac{\xi^2}{2\mu}-b_{\tau}\xi v+\mu b_{\tau}^2v^2.
    \nonumber
\end{equation}
Additionally, we could replace the latter result in (\ref{AppWKA6}) which returns
\begin{equation}
    \psi_{2}=\frac{b_{\tau}(2-\gamma\tau^2/m)}{4}\xi^2+\frac{b_{\tau}}{2}f_{\tau}(x,v,0)\xi+r_{2}(x,v).
    \nonumber
\end{equation}
Now we determine $r_{2}$ by invoking the next leading order:
\begin{align}
   4\mu\psi_{2\xi}^2-4(\xi-2\mu\psi_{1\xi})\psi_{3\xi}+f_{\tau}(x,v,\xi)\psi_{2v}+4v\psi_{2x}=0.
     \label{AppWKA9}
\end{align}
By repeating the procedure illustrate above once replaced the solution for $\psi_{2}$ in (\ref{AppWKA9}), we obtain according to \cite{kosek19881},
\begin{align}
    \frac{dr_{2}}{dt}&=\mu b_{\tau}^2\bigg(\frac{\mu a_{\tau}v}{4}(\gamma\tau^2/m-2)^2f_{\tau}(x,v,\xi_{0})-\frac{1}{4}f_{\tau}(x,v,\xi_{0})^2-v^2U_{xx}(\gamma\tau^2/m-2)\bigg) \nonumber \\
    &=\mu b_{\tau}^2\bigg(\frac{\mu a_{\tau}v}{4}\frac{d v}{dt}(\gamma\tau^2/m-2)^2-\frac{1}{2}(\gamma\tau^2/m-2)\bigg(U_{x}\frac{d v}{dt}+v^2U_{xx}\bigg)-\frac{1}{2}v^2U_{xx}(\gamma\tau^2/m-2)\bigg)\\
    &=\mu b_{\tau}^2\bigg(\frac{\mu a_{\tau}v}{4}\frac{d v}{dt}(\gamma\tau^2/m-2)^2-\frac{1}{2}(\gamma\tau^2/m-2)\frac{d}{dt}(v U_{x})-\frac{1}{2}vU_{xx}\bigg(\frac{4}{\mu a_{\tau}(\gamma\tau^2/m-2)}\frac{d v}{dt}-\frac{2}{\mu a_{\tau}}U_{x}\bigg)\bigg) \nonumber  
\end{align}
with  $\frac{dx}{dt}=v$ and $\frac{dv}{dt}=\frac{f_{\tau}(x,v,\xi_{0})}{2}$. By solving the above ordinary differential equation we get
\begin{equation}
    r_{2}(x,v)=\frac{\mu^2a_{\tau}b_{\tau}^2}{8}(\gamma\tau^2/m-2)^2v^2-\frac{\mu b_{\tau}^2}{2}(\gamma\tau^2/m-2)vU_{x}+\frac{b_{\tau}^2}{2a_{\tau}}(U_{x})^2-\frac{2b_{\tau}^2}{a_{\tau}(\gamma\tau^2/m-2)}\int_{0}^{t}v\frac{d v}{dt'}U_{xx}dt',
    \label{AppWKA9a}
\end{equation}
Substituting the result in (\ref{AppWKA2a}), we finally get the leading order solution up to $\mathcal{O}(\nu^2)$, i.e.
\begin{align}
    \phi_{0}&=a_{\tau}\frac{v^2}{2}+b_{\tau}U_{0}(x)+\frac{\tau_{an}}{m}\bigg( \frac{\xi^2}{2\mu}-b_{\tau}\xi v+\mu b_{\tau}^2v^2\bigg) +\Big(\frac{\tau_{an}}{m}\Big)^2\bigg(\frac{b_{\tau}\big(1-\frac{\gamma\tau^2}{2m}\big)}{2}\xi^2+\frac{b_{\tau}}{2}f_{\tau}(x,v,0)\xi \nonumber \\
    &+\frac{\mu^2b_{\tau}^3}{2}\bigg(1-\frac{\gamma\tau^2}{2m}\bigg)v^2+\mu b_{\tau}^2\bigg(1-\frac{\gamma\tau^2}{2m}\bigg)vU_{x}+\frac{b_{\tau}^2}{2a_{\tau}}(U_{x})^2+b_{\tau}\int_{0}^{t}v\frac{d v}{dt'}U_{xx}dt' \bigg).
    \label{AppSol0}
\end{align}
Now we go to the first-order in the perturbative parameter $D_{wn}$, which returns the expression,
\begin{align}
   \frac{4}{\tau_{an}}\bigg(\frac{2\mu}{\tau_{an}}\phi_{0\xi}-\xi\bigg)\phi_{1\xi}+\frac{\mu(2-\gamma\tau^2/m)}{\tau_{an}}\phi_{0v}^2+2f(x,v,\xi)\phi_{1v}+4v\phi_{1x}=0.
     \label{AppWKA10}
\end{align}
We can repeat the procedure illustrated previously to obtain an approximate solution for (\ref{AppWKA10}) in term of the perturbative parameter $\tau_{an}$. That is, inspired by the solution (\ref{InQPFH}) for the harmonic case, we propose the ansatz 
\begin{equation}
\phi_{1}=\psi_{-1}'\Big(\frac{\tau_{an}}{m}\Big)^{-1}+\psi_{0}'+\psi_{1}'\Big(\frac{\tau_{an}}{m}\Big)+\psi_{2}'\Big(\frac{\tau_{an}}{m}\Big)^2+\mathcal{O}(\nu^3)
\label{AppWKA2aSO}
\end{equation}
which returns recursive expressions for $\psi_{i}'$ for $i=-1,1,2,3$ after substituting it in (\ref{AppWKA10}). Concretely, we find 
\begin{align}
   \psi_{-1}'&=\frac{a_{\tau}}{2(\gamma\tau^2/m-2)}v^2-\frac{a_{\tau}}{2}U_{0}(x) \label{AppWKSO1}\\
    \psi_{0}'&=\frac{a_{\tau}}{2}v\xi-\mu a_{\tau}b_{\tau}v^2,  \\
    \psi_{1}'&=-\frac{b_{\tau}}{4}\xi^2-\frac{a_{\tau}}{4}h_{\tau}(x,v)\xi+r_{1}'(x,v),\label{AppWKSO2}
\end{align}
with $h_{\tau}(x,v)=6 v \gamma\tau + 3 v \mu (-2 +\gamma\tau^2/m) + (-2 + \gamma\tau^2/m) U_{x}$ and 
\begin{align}
    r_{1}'(x,v)&=\frac{3\mu^2a_{\tau}^3}{16}(\gamma\tau^2/m-2)^3v^2-\frac{\mu a_{\tau}^2}{4}(\gamma\tau^2/m-2)^2U_{x}(x)v\nonumber \\
    &-\frac{\mu a_{\tau}^2}{4}(\gamma\tau^2/m-2)^2\int_{0}^tv^2U_{xx}(x)dt',
    \nonumber
\end{align}
which was obtained from solving the following equation:
\begin{align}
    f_{\tau}(x,v,\xi_{0})r_{1v}'&+2vr_{1x}'=\frac{\mu^2a_{\tau}^2}{8}(\gamma\tau^2/m-2)^3\bigg(\frac{3a_{\tau}}{2}f_{\tau}(x,v,\xi_{0})v-\frac{2U_{xx}v^2}{\mu(\gamma\tau^2-2)} \bigg) \nonumber \\
    &+\frac{\mu a_{\tau}}{2}\big(f_{\tau}(x,v,0)\psi_{2\xi}(x,v,\xi_{0})+v(\gamma\tau^2/m-2)\psi_{2v}(x,v,\xi_{0}) \big) \nonumber \\
    &=\frac{\mu^2a_{\tau}^2}{8}(\gamma\tau^2/m-2)^3\bigg(\frac{3a_{\tau}}{2}f_{\tau}(x,v,\xi_{0})v-\frac{2U_{xx}v^2}{\mu(\gamma\tau^2/m-2)} \bigg) +\frac{\mu a_{\tau}}{2}\bigg(-\frac{a_{\tau}}{4}(\gamma\tau^2/m-2)f_{\tau}(x,v,0)f_{\tau}(x,v,\xi_{0})\nonumber \\
    &- \frac{a_{\tau}}{4}(\gamma\tau^2/m-2)v\bigg[ \frac{\mu a_{\tau}}{2}(\gamma\tau^2/m-2)^2f_{\tau}(x,v,\xi_{0})+2(\gamma\tau^2/m-2)\frac{\partial}{\partial v}\int_{0}^tU_{xx}v\frac{dv}{dt'}dt'\bigg]\bigg) \nonumber \\
    &=\frac{\mu^2a_{\tau}^2}{8}(\gamma\tau^2/m-2)^3\bigg(3a_{\tau}v\frac{dv}{dt}-\frac{2U_{xx}v^2}{\mu(\gamma\tau^2/m-2)} \bigg) -\frac{\mu a_{\tau}}{2}\bigg(\frac{a_{\tau}}{2}(\gamma\tau^2/m-2)f_{\tau}(x,v,0)\frac{dv}{dt}\nonumber \\
    &+\frac{a_{\tau}}{4}(\gamma\tau^2/m-2)v\bigg[\mu a_{\tau}(\gamma\tau^2/m-2)^2\frac{dv}{dt}+2(\gamma\tau^2/m-2)vU_{xx}\bigg]\bigg) \nonumber \\
    &=\frac{3\mu^2a_{\tau}^3}{8}(\gamma\tau^2/m-2)^3v\frac{dv}{dt}-\frac{\mu a_{\tau}^2}{4}(\gamma\tau^2/m-2)^2U_{x}\frac{dv}{dt}-\frac{\mu a_{\tau}^2}{2}(\gamma\tau^2/m-2)^2U_{xx}v^2 .
    \label{EqFODwnR2}
\end{align}
By substituting (\ref{AppWKSO1})-(\ref{AppWKSO2}) in (\ref{AppWKA2aSO}), we obtain the leading order solution of the quasipotential function, i.e.
\begin{align}
    \phi_{1}(x,v,\xi)&=-\frac{a_{\tau}}{2\tau_{an}}\bigg(\frac{v^2}{2\big(1-\frac{\gamma\tau^2}{2m}\big)}+U_{0}(x)\bigg)+ \frac{a_{\tau}}{2}v\xi-\mu a_{\tau}b_{\tau}v^2 \nonumber \\
    &-\Big(\frac{\tau_{an}}{m}\Big)\bigg( \frac{b_{\tau}}{4}\xi^2+\frac{a_{\tau}}{4}h_{\tau}(x,v)\xi+ \frac{3\mu^2b_{\tau}^3}{2}v^2+\mu b_{\tau}^2U_{x}(x)v\nonumber \\
    &+\mu b_{\tau}^2\int_{0}^tv^2U_{xx}dt'\bigg)+\mathcal{O}(\nu^2),
    \nonumber
\end{align}
which replaced in (\ref{AppWKBA}) together with (\ref{AppSol0}) yields the stationary probability distribution in the extended phase space $(X ,V,\xi_{an})$ approximated up to first order in the active-noise correlation time. Note that the last integral has no analytical solution of the bistable potential (\ref{PtFunc}), this can be approximated by numerical calculations. Recollecting these previous results together, we finally obtain the following expression for the generalized quasipotential function up to leading order in the perturbative parameters $\epsilon$, $\delta $, $\mu$ and $\tau$, i.e.
\begin{equation}
  \Psi(x,v,\xi)\approx\Psi_{wkb}^{(1)}(x,v)+\Psi_{wkb}^{(2)}(v,\xi),
  \label{AppWKBLSS}
\end{equation}
where we have introduced the auxiliary quadratic functions
\begin{align}
        \Psi_{wkb}^{(1)}(x,v)&=\frac{\mu}{4D_{an}}\bigg(a_{\tau}mv^2+2b_{\tau}U_{0}(x)+2\mu( b_{\tau}v)^2\tau_{an}\bigg)-\bigg[     \frac{a_{\tau}\mu}{8b_{\tau}D_{an}\tau_{an}}\bigg(a_{\tau}mv^2+2b_{\tau}U_{0}(x)+4\mu( b_{\tau}v)^2\tau_{an}\bigg)
            \nonumber \\
&+  \frac{b_{\tau}^2\mu^2\tau_{an}}{4m^2 
D_{an}}\bigg(3m\mu b_{\tau} v^2+2\bigg( m\Delta U^{-1}\int_{0}^t v^2U_{xx}dt'  +\sqrt{m\Delta U^{-1}}vU_{x}(x)\bigg)\bigg) \bigg]D_{wn},
\label{AppPSI1} \\
        \Psi_{wkb}^{(2)}(v,\xi)&=\frac{\tau_{an}}{2 }\bigg(\xi-\frac{2\mu b_{\tau}v}{\sqrt{2 D_{an}}} \bigg)\xi+\bigg[ \frac{\mu a_{\tau}v \xi}{2\sqrt{2 D_{an}}}  -\frac{\mu \tau_{an}\xi }{4}\bigg(\frac{b_{\tau}\xi}{m}-b_{\tau}\sqrt{{\frac{2}{D_{an}}}}\Big(3b_{\tau}m^{-1}v+(m^3\Delta U)^{-1/2}U_{x}(x)\Big)   \bigg)\bigg] D_{wn}.
    \label{AppPSI2}
\end{align}
Unfortunately, this expression get involve due to it contains an integral expression that must be evaluated by solving the non-linear dynamics evolution. As we are interested in an approximate description settled by the subsidiary conditions (\ref{PerWN_1}), (\ref{PerWN_2}) and (\ref{PerWN_3}), we shall consider the leading order of Eqs. (\ref{AppPSI1}) and (\ref{AppPSI2}) in the following; in particular, these will be employed in the derivation of the smallest non-vanishing eigenvalue in App. \ref{App4}. 

Lastly, we can determine the boundary layer function $K(x,v,\xi)=K_{0}(x,v,\xi)+\tau_{an}K_{1}(x,v,\xi)+\mathcal{O}(\nu^2)$ by carrying out the procedure described above to the next order in $\epsilon$: we find out at leading order in $D_{wn}$ that 
\begin{equation}
-\frac{2 (\xi-2\mu \tau_{an}^{-1}\phi_{0\xi})K_{\xi} }{\tau
   _{an}}+f_{\tau}(x,v,\xi)K_{v} +2 v K_{x}=0,
  \nonumber
\end{equation}
which lead us up to leading order in $\nu$ (or equivalently, up to leading order in $\tau_{an}$)
\begin{equation}
    2 (\xi-2\mu b_{\tau}v)K_{0\xi}=0,
    \nonumber
\end{equation}
which directly implies that $K_{0\xi}=0$ (as $\xi$ may take random values for which $\xi-2\mu b_{\tau}v\neq 0$) and thus, up to next order we get
\begin{equation}
2(\xi-2\mu b_{\tau}v)K_{1\xi} +f_{\tau}(x,v,\xi)K_{0v} +2 v K_{0x}=0.
  \nonumber
\end{equation}
Similarly as before in Eq. (\ref{AppWKA5}), we now impose that $K_{1}$ is non-singular in $\xi=2\mu b_{\tau}v $, which indirectly implies that $K_0$ is a constant function on any trajectory at leading order in $D_{wn}$ (by means of the characteristic method previously shown). Hence, our perturpative procedure returns that the boundary layer function is nearly constant, i.e. 
\begin{equation}
 K(x,v,\xi)=K_{0}, 
 \label{AppWKBLSSK}
\end{equation}
up to $\mathcal{O}(\epsilon^2)$, $\mathcal{O}(1)$ in $\delta$, and $\mathcal{O}(\nu^3)$. This result is consistent with similar treatments dealing with the Kramers equation \cite{naeh19901}. Notice that the reduced probability density (\ref{AppWKBASS}) and the quasipotential function (\ref{QPDSD}) provided in Sec. \ref{SubsectSSPDF} follow directly from Eq. (\ref{AppWKBA}) by integration over $\xi$ once we replace Eqs. (\ref{AppWKBLSS}) and (\ref{AppWKBLSSK}).

\section{Computation of the susceptibility with the linear response theory}\label{App4}
In this section we illustrate the computation of the escape rate as well as the derivation of the nonequilibrium stationary fluctuation-dissipation relation. These results are based on the steady-state probability distribution obtained in Sec. \ref{App2} in the weak noise limit. 

\subsection{Smallest non-vanishing eigenvalue in the weak noise limit}
Here we follow the flux-over-population method illustrated in Refs. \cite{kramers19401, schuller20201,naeh19901,hanggi19901} to compute the celebrated Kramers or escape rate $k_{ac}$ in the intermediate-to-strong friction regimen by means of the well-known flux-over-population method (since the potential is symmetric, the escape rate coincides). Accordingly, this is given by \cite{kramers19401}
\begin{equation}
    k_{ac}=\frac{j}{n_{x}},
\end{equation}
where $n_{x}$ and $j$ are the population in the initial well located at the metastable state $x=X^{\ast}_{1}$ and the current over the top of the potential barrier, respectively,
\begin{align}
    n_{x}=&\int_{-\infty}^{\infty}\int_{-\infty}^{\infty}\int_{-\infty}^{\infty}P(x,v,\xi)dxdvd\xi, \label{AppPNU2} \\
    j=&\int_{-\infty}^{\infty}\int_{-\infty}^{\infty}vP(X_{0}^{\ast},v,\xi)dvd\xi .\label{AppPNU2J}
\end{align}
By paying attention to these equations the escape rate is fully determined in the asymptotic time limit (i.e. $t\rightarrow\infty$) once we have obtained the steady-state solution $P_{st}(x,v, \xi)$ of the FP equation which satisfies the absorbing boundary condition at the point $X_{0}^{*}$ separating the two metastable states \cite{hanggi19901}. By invoking the Kramers ansatz, this can be written as 
\begin{equation}
    P_{st}(x,v, \xi)=Z^{-1}K(x,v, \xi) e^{-\Psi_{wkb}(x,v, \xi)},
    \label{AppKA1}
\end{equation}
where $\Psi_{wkb}(x,v, \xi)$ is the generalized quasipotential function previously obtained in App. \ref{App2}, which is given by Eq. (\ref{AppWKBLSS}), while $K(x,v, \xi)$ must satisfy the aforementioned absorbing boundary condition, that is
\begin{equation}
  K(x,v,\xi) =0, \ \ X>X_{0}^{*},
  \label{AppKDE}
\end{equation}
in contrast to the boundary layer function derived in App. \ref{App2}. However, to be consistent with the previous treatment we further require $K(x,v, \xi)$ to become the identity near the stable equilibrium states, this is known as the matching condition \cite{naeh19901}. According to escape rate theory, it is expected that the steady-state of the particle is well described by the probability density function  (\ref{AppWKBA}), obtained in the previous appendix, around the bottom wells \cite{hanggi19901}, which implies that $K(x,v, \xi)\approx 1$ with $x$ in a small area around the metastable states $X^{\ast}_{1,2}$, whereas $P_{st}(x,v, \xi)\approx 0$ for $x>X_{0}^{*}$, so that $K(x,v, \xi)\approx 0$. 

First, we obtain the stationary solution of Eq. (\ref{FPEq}) in the vanishing driving scenario (i.e., $F=0$) around the unstable point $X_{0}^{*}$, so that we could approximate the bistable potential as $U_{0}(x)\approx -\frac{1}{2}(\alpha-\gamma)(x-X^{\ast}_{0})^2$ in the small time-delay limit, and then, we substitute the Kramers' ansatz~(\ref{AppKA1}). This yields up to leading order (i.e., $\mathcal{O}(\tau_{an}^2,D_{wn}\tau_{an})$)
\begin{align}
  &\Bigg[-v\frac{\partial}{\partial x}-\frac{b_{\tau}}{ma_{\tau}}\bigg((\alpha-\gamma)(x-x^{\ast}_{0})-b_{\tau}\mu v+\sqrt{2D_{an}}\xi+\frac{\mu D_{wn}}{m\tau_{an}}\bigg(a_{\tau}m v+b_{\tau}\tau_{an}\big( 2b_{\tau}\mu v-\sqrt{2D_{an}}\xi \big)\bigg)\bigg)\frac{\partial}{\partial v} \nonumber \\
  &+\Bigg(\frac{
a_{\tau}m
\left(
\sqrt{2}\,b_{\tau}v\mu
-\sqrt{D_{an}}\,\xi
\right)
+
b_{\tau}^{2}\mu\tau_{an}
\left(
\sqrt{2}\,x(-\alpha+\gamma)
+\sqrt{2}\,b_{\tau}v\mu
-2\sqrt{D_{an}}\,\xi
\right)
}
{a_{\tau}\sqrt{D_{an}}\,m\,\tau_{an}} +\frac{\mu D_{wn}}{2a_{\tau}\sqrt{D_{an}}\,m^{2}\tau_{an}^{2}}
\Big(
-\sqrt{2}\,a_{\tau}^{2}m^{2}v  \nonumber \\
&+b_{\tau}^{3}\mu\tau_{an}^{2}
\left(
\sqrt{2}\,x(\alpha-\gamma)
-3\sqrt{2}\,b_{\tau}v\mu
+4\sqrt{D_{an}}\,\xi
\right)
+a_{\tau}b_{\tau}m\tau_{an}
\left(
-3\sqrt{2}\,b_{\tau}v\mu
+2\sqrt{D_{an}}\,\xi
+\sqrt{2}(\alpha-\gamma)
(x-v\tau_{an})
\right)
\Big)
\Bigg)\frac{\partial}{\partial \xi}  \nonumber\\
&+ \frac{b_{\tau}D_{an}D_{wn}}{a_{\tau}m^2\tau_{an}}\frac{\partial^2}{\partial v^2}+\frac{1}{\tau_{an}^2}\frac{\partial^2}{\partial \xi^2}\Bigg]K(x,v,\xi)=0, \nonumber
\end{align}
where we have made use of the condition (\ref{AppKDE}). The second important assumption in the Kramers derivation consists of assuming that $K(x,v, \xi)$ is a function of a linear combination of position and velocity \cite{kramers19401}. Here, we generalize the latter to contain the active noise as follows
\begin{equation}
  K(x,v, \xi)=K(v-A_{\tau}\xi-B_{\tau}(x-X^{\ast}_{0}))=K(z),
   \label{AppKA2A}
\end{equation}
where the coefficients $A_{\tau}$ and $B_{\tau}$ must be determined. The ansatz (\ref{AppKA2A}) leads to a one-dimensional parabolic differential equation, that is
\begin{align}
 \Bigg[
&\frac{1}{a_{\tau}\sqrt{D_{an}}\,m\tau_{an}}
\Bigg(
\sqrt{D_{an}}
\left(
a_{\tau}bmv
+b_{\tau}
\left(
-x\alpha+x\gamma+b_{\tau}v\mu
-\sqrt{2}\sqrt{D_{an}}\,\xi
\right)
\right)\tau_{an}
\nonumber\\
&\qquad
+A_{\tau}\left(
a_{\tau}m
\left(
-\sqrt{2}b_{\tau}v\mu
+\sqrt{D_{an}}\,\xi
\right)
+b_{\tau}^{2}\mu
\left(
\sqrt{2}x(\alpha-\gamma)
-\sqrt{2}b_{\tau}v\mu
+2\sqrt{D_{an}}\,\xi
\right)\tau_{an}
\right)
\nonumber\\
&\qquad
+\frac{\mu
\left(
-2b_{\tau}\sqrt{D_{an}}\,\tau_{an}
\left(
a_{\tau}mv
+b_{\tau}
\left(
2b_{\tau}v\mu
-\sqrt{2}\sqrt{D_{an}}\,\xi
\right)\tau_{an}
\right)
\right)}
{2m\tau_{an}}
\nonumber\\
&\qquad
+A_{\tau}\left(
\sqrt{2}\,a_{\tau}^{2}m^{2}v
+b_{\tau}^{3}\mu
\left(
\sqrt{2}x(-\alpha+\gamma)
+3\sqrt{2}\,b_{\tau}v\mu
-4\sqrt{D_{an}}\,\xi
\right)\tau_{an}^{2}
\right.
\nonumber\\
&\qquad\qquad\left.
+a_{\tau}b_{\tau}m\tau_{an}
\left(
3\sqrt{2}\,b_{\tau}v\mu
-2\sqrt{D_{an}}\,\xi
-\sqrt{2}(\alpha-\gamma)(x-v\tau_{an})
\right)
\right)D_{wn}
\Bigg)
\frac{\partial}{\partial z}
\nonumber\\
&+
\left(
\frac{a_{\tau}^{2}}{\tau_{an}^{2}}
+
\frac{b_{\tau}D_{an}D_{wn}}
{a_{\tau}m^{2}\tau_{an}}
\right)
\frac{\partial^{2}}{\partial z^{2}}
\Bigg]K(z)=0.
\label{AppKA2Eq}
\end{align}
by doing the replacement
\begin{align}
 &\frac{1}{a_{\tau}\sqrt{D_{an}}\,m\tau_{an}}
\Bigg(
\sqrt{D_{an}}
\left(
a_{\tau}bmv
+b_{\tau}
\left(
-x\alpha+x\gamma+b_{\tau}v\mu
-\sqrt{2}\sqrt{D_{an}}\,\xi
\right)
\right)\tau_{an}
\nonumber\\
&\qquad
+A_{\tau}\left(
a_{\tau}m
\left(
-\sqrt{2}b_{\tau}v\mu
+\sqrt{D_{an}}\,\xi
\right)
+b_{\tau}^{2}\mu
\left(
\sqrt{2}x(\alpha-\gamma)
-\sqrt{2}b_{\tau}v\mu
+2\sqrt{D_{an}}\,\xi
\right)\tau_{an}
\right)
\nonumber\\
&\qquad
+\frac{\mu
\left(
-2b_{\tau}\sqrt{D_{an}}\,\tau_{an}
\left(
a_{\tau}mv
+b_{\tau}
\left(
2b_{\tau}v\mu
-\sqrt{2}\sqrt{D_{an}}\,\xi
\right)\tau_{an}
\right)
\right)}
{2m\tau_{an}}
\nonumber\\
&\qquad
+A_{\tau}\left(
\sqrt{2}\,a_{\tau}^{2}m^{2}v
+b_{\tau}^{3}\mu
\left(
\sqrt{2}x(-\alpha+\gamma)
+3\sqrt{2}\,b_{\tau}v\mu
-4\sqrt{D_{an}}\,\xi
\right)\tau_{an}^{2}
\right.
\nonumber\\
&\qquad\qquad\left.
+a_{\tau}b_{\tau}m\tau_{an}
\left(
3\sqrt{2}\,b_{\tau}v\mu
-2\sqrt{D_{an}}\,\xi
-\sqrt{2}(\alpha-\gamma)(x-v\tau_{an})
\right)
\right)D_{wn}
\Bigg)=\lambda_{\tau} z =\lambda_{\tau}(v-A_{\tau}\xi-B_{\tau}(x-x^{\ast}_{0})). \label{AppKA2}
\end{align}
Paying attention to coefficients in (\ref{AppKA2}), we get 
\begin{align}
 B_{\tau}&= \frac{
b_{\tau}(\alpha-\gamma)
\left(
2\sqrt{D_{an}}\,m\tau_{an}
+
\sqrt{2}\,a\mu
\left(
a_{\tau}D_{wn}m
+
b_{\tau}
\left(
-2m
+
b_{\tau}D_{wn}\mu
\right)
\tau_{an}
\right)
\right)
}
{2a_{\tau}\sqrt{D_{an}}\,m^{2}\lambda_{\tau}\tau_{an}}, \label{KRB} \\
   A_{\tau}&=\frac{
\sqrt{2}\,b_{\tau}\sqrt{D_{an}}
\left(
-m+b_{\tau}D_{wn}\mu
\right)
\tau_{an}
}
{
2b_{\tau}^{2}\mu
\left(
-m+b_{\tau}D_{wn}\mu
\right)
\tau_{an}
-
a_{\tau}m
\left(
m-b_{\tau}D_{wn}\mu
+m\lambda_{\tau}\tau_{an}
\right)
}, \label{KRA}
\end{align}
which yields the following quadratic equation for $\lambda$ up to leading order,
\begin{equation}
   a_{\tau}m\lambda_{\tau}^{2}
+
b_{\tau}
\left(
-\alpha+\gamma+b_{\tau}\lambda_{\tau}\mu
\right)=0, \nonumber
\end{equation}
whose roots are
\begin{align}
  \lambda^{\pm}_{\tau}=  \frac{
-b_{\tau}^{2}\mu
+
\sqrt{b_{\tau}}\,
\sqrt{
4a_{\tau}m(\alpha-\gamma)
+b_{\tau}^{3}\mu^{2}
}
}
{2a_{\tau}m}. 
  \label{AppEqRoot}
\end{align}
This coincides with the result obtained for the white-noise scenario.  Replacing this result in the above expressions, we obtain $A_{\tau}^{\pm}$ and $B_{\tau}^{\pm}$ (i.e., by substituting $\lambda_{\tau}\rightarrow\lambda^{\pm}_{\tau}$ in (\ref{KRA}) and (\ref{KRB})). Going back to the solution of Eq. (\ref{AppKA2Eq}), this is given by 
\begin{equation}
    K(z)=\sqrt{\frac{\lambda_{+}}{2\pi C_{+}}}\int_{-\infty}^{z}\exp\bigg(-\frac{\lambda_{+}s^2}{2C_{+}}\bigg)d s=\frac{1}{2}\text{erf}(z\sqrt{c_{+}}),
    \label{AppSK}
\end{equation}
where we have chosen $\lambda=\lambda_{\tau}^{+}$ to guarantee the convergence of this solution \cite{schuller20201}, and introduced the parameters $c^{+}_{\tau}=\frac{\lambda^{+}_{\tau}}{2C^{+}_{\tau}}$ and   
\begin{equation}
   C^{+}_{\tau}=\frac{
(A_{\tau}^{+})^2+\dfrac{b_{\tau}D_{an}D_{wn}\tau_{an}}
{a_{\tau}m^2}
}
{\tau_{an}^2}.
    \nonumber
\end{equation}
Now, by following Kramers procedure, we compute the population around the metastable state $X^{\ast}_{1}$ by approximating the potential, i.e. $U_{0}(x)\approx-\Delta U+(\alpha-\gamma)(x-X^{\ast}_{1})^2$, which returns
\begin{align}
    n_{x}&=\frac{1}{Z}\int_{-\infty}^{\infty}\int_{-\infty}^{\infty}\int_{-\infty}^{\infty}e^{-\Psi_{wkb}(x,v,\xi)} d \xi d v dx , \nonumber \\
    &=\frac{1}{Z}\sqrt{\frac{2\pi}{\tau_{an}}}\int_{-\infty}^{\infty}\int_{-\infty}^{\infty}e^{-\Phi_{wkb}(x,v)}d v dx\nonumber \\
    &\approx\frac{e^{d_{\tau}\Delta U}}{Z}\sqrt{\frac{2\pi}{\tau_{an}}}\int_{-\infty}^{\infty}\int_{-\infty}^{\infty}e^{-c_{\tau} v^2-d_{\tau}(\alpha-\gamma)(x-x_{1}^{\ast})^2} d v dx=\frac{e^{d_{\tau}\Delta U}}{Z}\sqrt{\frac{2\pi^3}{\tau_{an}c_{\tau}d_{\tau}(\alpha-\gamma)}}, \nonumber
\end{align}
where we have assumed $\tau_{an}>0$ and substituted the expression for the generalized quasipotential function (\ref{AppWKBLSS}) obtained in App. \ref{App2} at leading order in the perturbative parameters. Similarly, by replacing (\ref{AppSK}) into (\ref{AppPNU2J}), we compute the current over the barrier top (this means $x=X^{\ast}_{0}=0$) by taking advantage of the standard tables of integration for error functions \cite{ng19691}, this yields
\begin{align}
    j&=\frac{1}{2Z}\int_{-\infty}^{\infty}\int_{-\infty}^{\infty}ve^{-\Psi_{wkb}(X^{\ast}_{0},v,\xi)}\text{erf}\big((v+A^{+}_{\tau}\xi)\sqrt{c^{+}_{\tau}}\big) d \xi d v , \nonumber \\
    &=\frac{Z^{-1}}{2A^{+}_{\tau}\sqrt{{c^{+}_{\tau}}}}\int_{-\infty}^{\infty}\int_{-\infty}^{\infty}ve^{-\Psi^{(2)}_{wkb}(X^{\ast}_{0},v)-\Psi_{wkb}^{(1)}\big(v,\frac{z}{A^{+}_{\tau}\sqrt{{c^{+}_{\tau}}}}-\frac{v}{A^{+}_{\tau}}\big)}\text{erf}(z) d z d v\nonumber \\
    &=-\sqrt{\frac{\pi Z^{-2}}{4(A^{+}_{\tau}p_{\tau})^2c^{+}_{\tau}}}e^{-d_{\tau}U_{0}(X^{\ast}_{0})}\int_{-\infty}^{\infty}ve^{-e_{\tau}v^2}\text{erf}\bigg(\frac{q_{\tau}v }{\sqrt{p_{\tau}^2+1}}\bigg)dv \nonumber \\
    &=-\sqrt{\frac{\pi}{c^{+}_{\tau}}}\frac{\frac{q }{\sqrt{p^2+1}}}{2A^{+}_{\tau}p_{\tau} Ze_{\tau}}\bigg(\frac{q_{\tau}^2}{p_{\tau}^2+1}+e_{\tau}\bigg)^{-\frac{1}{2}}e^{-d_{\tau}U_{0}(x^{\ast}_{0})}\nonumber \\
    &\approx\frac{-qe^{-d_{\tau}U_{0}(X^{\ast}_{0})}}{2A^{+}_{\tau}p_{\tau}Ze_{\tau}
    }\sqrt{\frac{\pi}{c^{+}_{\tau}(q_{\tau}^2+e_{\tau}(p_{\tau}^2+1))}},
    \nonumber
\end{align}
where we have introduced the coefficients,
\begin{align}
    p_{\tau}&=\sqrt{\frac{
\tau_{an}
\left(
a_{\tau}m+b_{\tau}^{2}\mu\tau_{an}
\right)
}
{2a_{+}^{2}a_{\tau}c_{+} m}
-
\frac{
b_{\tau}\mu\tau_{an}
\left(
a_{\tau}m+2b_{\tau}^{2}\mu\tau_{an}
\right)
D_{wn}
}
{4a_{+}^{2}a_{\tau}c_{+} m^{2}}}  ,\nonumber \\
    q_{\tau}&=-\frac{a_{\tau}\,\mu D_{wn}
}
{4\sqrt{D_{an}\tau_{an}}}+\frac{
\left[
4a_p b_{\tau}\mu
\left(
m-b_{\tau}D_{wn}\mu
\right)
+
\sqrt{2}\sqrt{D_{an}}
\left(
-4m+b_{\tau}D_{wn}\mu
\right)
\right]
\sqrt{\tau_{an}}
}
{8a_p\sqrt{D_{an}}\,m},
    \nonumber \\
    e_{\tau}&=-\frac{
a_{\tau}^{2}D_{wn}m\mu
}
{8b_{\tau}D_{an}\tau_{an}}+\frac{
\mu
\left(
a_{\tau}
\left(
m-b_{\tau}D_{wn}\mu
\right)
+
b_{\tau}^{2}\mu\tau_{an}
\right)
}
{4D_{an}}. \nonumber
\end{align}
A Taylor series expansion for sufficiently small noise strength and active noise correlation time returns 
\begin{align}
 k=  \frac{\lambda^{+}_{\tau}}{\sqrt{2}\pi}\bigg(1-\frac{b_{\tau}^2}{2a_{\tau}}\frac{\mu \tau_{an}}{m} \bigg)e^{-\frac{\mu \Delta U_{\mathrm{eff}}}{2D_{an}}}+\mathcal{O}(\tau_{an}^2,D_{wn}\tau_{an},\gamma\tau^3), \label{AppEcLEV} 
\end{align}
where $\Delta U_{\mathrm{eff}}$ is the renormalized potential barrier height given by Eq.~(\ref{EqPBEU}). Equation (\ref{MFRT_1}) follows from (\ref{AppEcLEV}) after replacing (\ref{AppEqRoot}) and some algebra. One may also obtain the escape rate in the overdamped scenario from (\ref{MFRT_1}) by replacing $\lambda^{+}_0\rightarrow(\alpha-\gamma)/\mu(1+\gamma\tau/\mu)$, i.e.
\begin{align}
    k_{ac}&= \frac{\alpha-\gamma}{\sqrt{2}\pi\mu}\Bigg[1+\frac{\gamma\tau}{\mu}-\frac{\mu\tau_{an}}{2m} \Bigg]e^{-\frac{(\mu-\gamma\tau)\Delta U }{2 D_{an}}\big(1-\frac{D_{wn}}{2\tau_{an}} \big)}+\mathcal{O}\bigg(\frac{(\gamma\tau)^2}{\mu^2},\frac{(\mu\tau_{an})^2}{m^2},\frac{D_{wn}^2}{\tau_{an}^2}\bigg). \label{MFRT_OD}
\end{align}

Owing to the clear-cut separation between the interwell dynamics and the intrawell dynamics, the Kramers rate can be asymptotically identified with the smallest (real part of) eigenvalue $\lambda_{ac}$ in the small noise limit \cite{hanggi19861,holcman20141}, as stated in Sec. \ref{SubsectSSPDF}. From above expression we finally obtain an estimation of the latter \cite{jung19931,hanggi19901}, i.e. $\lambda_{ac}\approx 2j/n_{x}$, 
\begin{equation}
    \lambda_{ac}\approx\frac{\sqrt{2}\lambda^{+}_{\tau}}{\pi}\bigg(1-\frac{b_{\tau}^2}{2a_{\tau}}\frac{\mu \tau_{an}}{m} \bigg)e^{-\frac{\mu \Delta U_{\mathrm{eff}}}{2D_{an}}}.\label{AppEcLEING} 
\end{equation}
Now one may get an estimation of the SPA from Eq. (\ref{AppEcLEING}), once substituted the expression for the susceptibility (\ref{EqSuc}), which yields
\begin{align}
    \eta_{an}(\tau,\tau_{an}, D_{wn},D_{an})&\approx\frac{a_{\tau}m\mu^2\Delta U}{b_{\tau}\beta D_{an}^2}\bigg(b_{\tau}\big(m a_{\tau}+2b_{\tau}^2\mu\tau_{an}\big)-ma_{\tau}^2\frac{D_{wn}}{\tau_{an}}  \bigg) \nonumber \\ 
    &\times \bigg[1+\bigg(\frac{\pi\Omega}{\sqrt{2}\lambda^{+}_{\tau}}\bigg)^2 \bigg(1+\frac{b_{\tau}^2\mu \tau_{an}}{a_{\tau}m}  \bigg)e^{\frac{\mu \Delta U_{\mathrm{eff}}}{D_{an}}}\bigg]^{-1},
    \nonumber
\end{align}
After taking the leading order in the small time-delay from the above expression, i.e. $a_{\tau}\rightarrow(1-\gamma\tau/\mu)$ and $b_{\tau}\rightarrow a_{\tau}$, the SPA can be cast in the following form
\begin{align}
  \eta_{ac}(\tau,\tau_{an}, D_{wn},D_{an}) &\approx \frac{m^2 \mu^2\Delta U}{4\beta D_{an}^2}\bigg(1-2\frac{\gamma\tau}{\mu}\bigg)\Bigg(1+2 \frac{\mu\tau_{an}}{m}\bigg(1-\frac{\gamma\tau}{\mu} \bigg)-\frac{D_{wn}}{\tau_{an}} \bigg)\Bigg)\nonumber \\
&\times\Bigg[1+\Bigg(\bigg(\frac{\pi\Omega}{\sqrt{2}\lambda^{+}_{0}}\bigg)^2 -\frac{\sqrt{2}\pi\Omega}{\lambda^{+}_{0}-2(\alpha-\gamma)\mu^{-1}}\frac{\gamma\tau}{\mu}\Bigg)\Bigg(1+\bigg(1-\frac{\gamma\tau}{\mu}\bigg)\frac{\mu \tau_{an}}{m}  \Bigg)e^{\frac{\mu \Delta U}{D_{an}}\big(1-\frac{\gamma\tau}{\mu}\big)\big(1-\frac{D_{wn}}{2\tau_{an}}\big)}\Bigg]^{-1}.
  \label{etaLRT}
\end{align}
Equation~(\ref{etaLRTapp}) follows from (\ref{etaLRT}) after explicitly introducing the result for the purely white-noise case, corresponding to the passive bistable particle scenario, namely $\eta_{pass}$.

\subsection{Approximate response function}

In this section, we briefly illustrate the linear response theory and the derivation of the approximated expression provided by Eq.~(\ref{EqSuc}), following the prescription stated in \cite{jung19931,jung19911,hanggi19931}. 

Accordingly, for a sufficiently small driving frequency and
amplitude relative to the potential-well scale and the active-noise
intensity (see conditions (\ref{EqF1})--(\ref{EqF3})), the asymptotic
average value of $X(t)$ can be obtained from
\cite{hanggi19821,hanggi19931,jung19931,jung19932}
\begin{equation}
    \langle X(t)\rangle
    =\frac{F}{m}\int_{-\infty}^{\infty}R(t-t')\cos(\Omega t')dt',
\end{equation}
where $R(t)$ denotes the response function \cite{jung19931}, i.e., with $t>t_{0}$
\begin{align} 
     R(t-t_{0})&=\int_{-\infty}^{+\infty}\int_{-\infty}^{+\infty}x_{0} e^{\mathbb{L}_{FP}(t-t_{0})}g(x_{0},v_{0})P_{st}(x_{0},v_{0})dx_{0}dv_{0}  \nonumber \\
     &=-\int_{-\infty}^{+\infty}\int_{-\infty}^{+\infty}x_{0} e^{\mathbb{L}_{FP}(t-t_{0})}\frac{\partial}{\partial v_{0}}P_{st}(x_{0},v_{0})dx_{0}dv_{0},
     \label{AppKKR2}
\end{align}
where $g(x_{0},v_{0})=(P_{st}(x_{0},v_{0}))^{-1}\big(-\frac{\partial }{\partial v_{0}}P_{st}(x_{0},v_{0})\big)$, $\mathbf{L}_{FP}$ is the FP operator retrieved by Eq. (\ref{FPEq}), and recall that $P_{st}(x,v)$ denotes the stationary probability of the unperturbed system (i.e., $F=0$).

Equivalently, linear-response theory implies that the mean response remains
harmonic at the driving frequency
\cite{hanggi19901,dykman19931,hanggi19931,anishchenko19991}, i.e.,
\begin{equation}
    \langle X(t)\rangle_{as}
    =2|\tilde{X}_{1}|\cos(\Omega t+\bar{\phi}_{1}),
    \nonumber
\end{equation}
where the amplitude $\bar{X}_{1}$ and the phase lag $\bar{\phi}_{1}$
of the first moment of the spatial coordinate at the driving frequency
$\Omega$ are given by
\begin{align}
    |\tilde{X}_{1}|&=\frac{F}{2m}|\chi(\Omega)|,
    \label{FTRFS1app} \\
    \bar{\phi}_{1}&=-\arctan\bigg(
    \frac{\text{Im}\,\chi(\Omega)}{\text{Re}\,\chi(\Omega)}
    \bigg),
    \nonumber
\end{align}
and $\chi(\omega)$ denotes the Fourier transform of the response function
$R(t)$, i.e., the susceptibility,

\begin{equation}
    \chi(\omega)=\int_{-\infty}^{+\infty}e^{-i\omega t}R(t)dt.
    \label{FTRFS2}
\end{equation}

Now we show that the response function can be recast into the following form 
\begin{equation}
    R(t)=-2c_{\tau}\frac{d}{dt}K_{xx}(t),
    \label{AppKKR1}
\end{equation}
where $c_{\tau}$ is given by Eq. (\ref{AppCWKB1}) in Sec. \ref{SubsectSSPDF}, and $K_{xx}(t)=\langle x(t_{0})x(t_{0}+t)\rangle_{as}$. Concretely, equation (\ref{AppKKR1}) is derived by appealing to the stationary property of the unperturbed asymptotic state \cite{risken19841}, i.e. $\langle v(t_{0})x(t_{0}+t)\rangle_{as}=-\langle x(t_{0})v(t_{0}+t)\rangle_{as}$, which yields
\begin{align}
\frac{d}{dt}K_{xx}(t)&= \langle x(t_{0})v(t_{0}+t)\rangle_{as} \nonumber \\
&=-\langle v(t_{0})x(t_{0}+t) \rangle_{as} \nonumber \\
&=-\int_{-\infty}^{+\infty}\int_{-\infty}^{+\infty}x_{0} e^{\mathbb{L}_{FP}(t-t_{0})}v_{0}P_{st}(x_{0},v_{0})dx_{0}dv_{0}.
    \nonumber
\end{align}
We further build upon the results obtained in App. \ref{App2} and consider $P_{st}(x,v)\approx Z^{-1}e^{-\Phi_{wkb}(x,v)}$ (recall that $\Phi_{wkb}(x,v)$ is given by Eq. (\ref{QPDSD})), which remains justified at leading order in $D_{an}$, $\tau$, $\tau_{an}$ and $D_{wn}$. Notice that $P_{st}(x,v)$ is exactly Maxwellian in the velocity, which immediately returns 
\begin{equation}
    \frac{\partial}{\partial v}P_{st}(x,v)=-2c_{\tau}vP_{st}(x,v).
    \nonumber
\end{equation}
Once we replaced the above result in Eq. (\ref{AppKKR2}), after some manipulation we arrive at the expression (\ref{AppKKR1}). Let us emphasize that the latter is a non-equilibrium steady-state fluctuation-response relation, valid without detailed balance, resting entirely on (i) the general identity (\ref{AppKKR2}) holding for a Markovian FP process, (ii) the stationary property of the asymptotic state of the FP operator, and (iii) the Maxwellian shape of $P_{st}$ established separately in App.~\ref{App2}. Next, we make use of the long-time approximation for the correlation function \cite{jung19931,hanggi19931,anishchenko19991}, that is 
\begin{equation}
    K_{xx}(t)\approx  \langle x ^2\rangle_{st}e^{-\lambda_{ac}t} \ \ \ \ t>0,
    \label{EqLTApp}
\end{equation}
where $\lambda_{ac}$ is approximately given by Eq. (\ref{EqAER}). Replacing the latter result in Eq. (\ref{AppKKR1}) leads to the approximate expression (\ref{EqSuc}) for the susceptibility once we perform the Fourier transform of Eq. (\ref{FTRFS2}), which is consistent within the domain of validity determined by the subsidiary conditions (\ref{PerTD}), (\ref{PerWN_1}), and (\ref{EqF1})-(\ref{EqF3}). Hence, to obtain an analytical expression for the SPA, we first substitute Eq.~(\ref{EqAER}) into Eq.~(\ref{EqSuc}) and insert the resulting solution into Eq.~(\ref{FTRFS1}). The corresponding Fourier response is then substituted into Eq.~(\ref{SPAE}), yielding the approximate expression for the SPA.

We also notice that $ \langle x ^2\rangle_{st}$ is computed from the stationary (unperturbed) probability density function (\ref{AppWKBASS}), we get
\begin{align}
     \langle x ^2\rangle_{st}&=\frac{1}{Z}\int_{-\infty}^{\infty}\int_{-\infty}^{\infty}x^2e^{-c_{\tau}v^2-d_{\tau}U_{0}(x)} dxdv\nonumber \\
     &=\frac{\pi}{2\sqrt{c_{\tau}}Z}\bigg(\frac{\beta d_{\tau}}{2}\bigg)^{-\frac{3}{4}}e^{\frac{(\alpha-\gamma)^2d_{\tau}}{8\beta}}D_{-\frac{3}{2}}\bigg(-\frac{(\alpha-\gamma)\sqrt{d_{\tau}}}{\sqrt{2\beta}} \bigg) \nonumber \\
     &=\sqrt{\frac{1}{2\beta d_{\tau}}}\frac{D_{-\frac{3}{2}}\big(-\frac{(\alpha-\gamma)\sqrt{d_{\tau}}}{\sqrt{2\beta}} \big)}{D_{-\frac{1}{2}}\big(-\frac{(\alpha-\gamma)\sqrt{d_{\tau}}}{\sqrt{2\beta}} \big)},
     \nonumber
\end{align}
where we have previously computed the normalizing factor
\begin{equation}
    Z=\frac{\pi}{\sqrt{c_{\tau}}}\bigg(\frac{\beta d_{\tau}}{2}\bigg)^{-\frac{1}{4}}e^{\frac{(\alpha-\gamma)^2d_{\tau}}{8\beta}}D_{-\frac{1}{2}}\bigg(-\frac{(\alpha-\gamma)\sqrt{d_{\tau}}}{\sqrt{2\beta}} \bigg), \nonumber
\end{equation}
with $D_{\nu}(z)$ denoting the parabolic cylinder function (with $\nu\in \mathbb{Z}$ and $z\in \mathbb{C}$). Let us recall that in the interesting stochastic scenario $1\ll\frac{(\alpha-\gamma)^2d_{\tau}}{8\beta}$, we may replace the asymptotic approximation of these with large arguments which yields
\begin{align}
     \langle x ^2\rangle_{st}\approx \frac{\alpha-\gamma}{\beta}-\frac{D_{an}}{b_{\tau}\mu(\alpha-\gamma)}\bigg[2+\frac{a_{\tau}D_{wn}}{b_{\tau}\tau_{an}}\bigg].
     \nonumber
\end{align}
We arrive at Eq. (\ref{EqASXS}) by keeping leading order terms once we substituted the coefficients (\ref{ACoefa}) and (\ref{ACoefb}), as previously explained. 

\section{Numerical convergence analysis}\label{App5}

\begin{figure}[h!]
\centering
\includegraphics[width=0.9\textwidth]{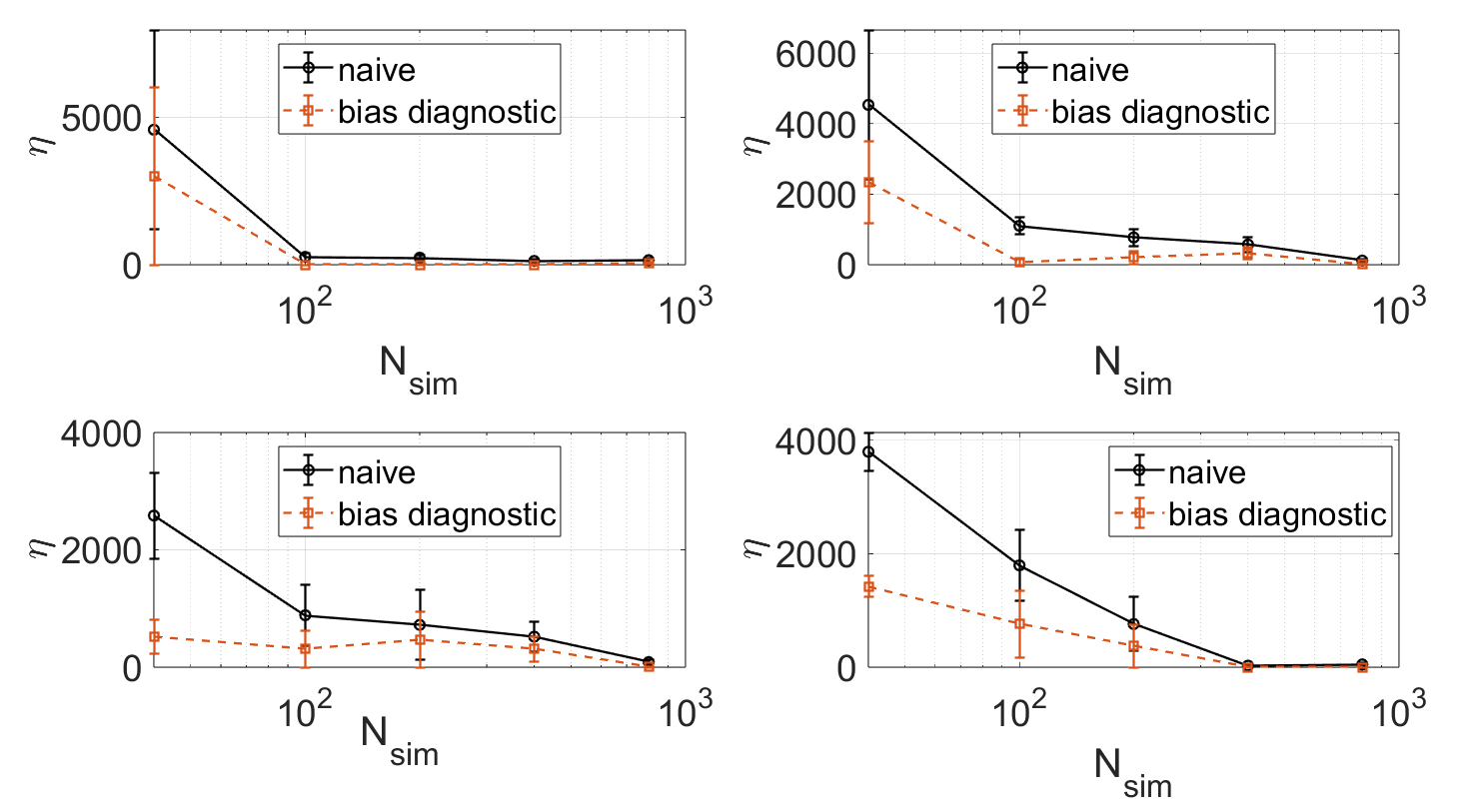} 
\caption{SPA as a function of the number $N_{sim}$ of trajectories composing the ensemble of numerical simulations. Convergence of the numerically estimated SPA with the ensemble size $N_{sim}$ for four representative points in the resonant region. The upper panels correspond to $D_{\mathrm{wn}}=0$, $\tau_{\mathrm{an}}=0.1$, and $\tau=0$, with $D_{\mathrm{an}}=20$ (left) and $D_{\mathrm{an}}=22.5$ (right). The lower panels correspond to $D_{\mathrm{wn}}=0.05$, $\tau_{\mathrm{an}}=0.1$, and $\tau=0$, with $D_{\mathrm{an}}=15$ (left) and $D_{\mathrm{an}}=17.5$ (right). The solid curves show the standard SPA estimator, whereas the dashed curves provide a finite-ensemble bias diagnostic obtained by subtracting the variance contribution associated with the residual incoherent Fourier component. Increasing $N_{sim}$ strongly reduces the apparent SPA amplitude, while the existence and location of the resonant region remain robust.}
\label{fig:CAS}
\end{figure}

To assess the statistical robustness of the SPA amplitude, we performed an additional convergence study at four representative points located in the resonant region. For each parameter set, the same numerical procedure was repeated using nested ensembles with $N_{sim}=40,100,200,400,$ and $800$ stochastic trajectories. Fig.~\ref{fig:CAS}  shows that the absolute SPA amplitude is strongly affected by finite-ensemble statistics: the large values obtained for small $N$ decrease substantially as the ensemble size is increased.

This behavior is consistent with the fact that the SPA is constructed from the squared modulus of a coherent Fourier average. For finite ensembles, the residual incoherent contribution to the complex Fourier coefficient does not vanish completely and therefore produces a positive finite-$N$ contribution to the estimated SPA. The dashed curves in Fig.~\ref{fig:CAS} provide a diagnostic estimate of this contribution and are included only to assess the importance of finite-ensemble effects; they are not used as an alternative definition of the SPA.

Importantly, although the absolute amplitude converges more slowly with the number of stochastic realizations, the non-monotonic response and the location of the resonant region are considerably more robust. This supports the use of the numerical SPA primarily to characterize the existence and approximate position of the stochastic-resonance region, whereas the matching condition obtained from the lowest non-vanishing eigenvalue provides a more quantitatively stable comparison with the analytical theory.

\bibliographystyle{apsrev4-1}
\bibliography{references}
\end{document}